\documentclass[%
 sor4-1,
 reprint,
 amsmath,amssymb,
]{revtex4-1}

\usepackage{graphicx}
\usepackage{dcolumn}
\usepackage{bm}
\usepackage{xcolor}
\usepackage{enumerate}

\begin{document}
	
\preprint{APS/123-QED}

\title{Medium Amplitude Parallel Superposition (MAPS) Rheology \\ Part 1: Mathematical Framework and Theoretical Examples}
\author{Kyle R. Lennon$^1$}
\author{Gareth H. McKinley$^2$}
\author{James W. Swan$^1$}
\thanks{Corresponding author; Electronic mail: jswan@mit.edu}
\affiliation{1. Department of Chemical Engineering, \\ 2. Department of Mechanical Engineering, Massachusetts Institute of Technology, Cambridge, MA \\}
\date{\today}

\begin{abstract}
A new mathematical representation for nonlinear viscoelasticity is presented based on application of the Volterra series expansion to the general nonlinear relationship between shear stress and shear strain history.  This theoretical and experimental framework, which we call Medium Amplitude Parallel Superposition (MAPS) Rheology, reveals a new material property, the third order complex modulus, which describes completely the weakly nonlinear response of a viscoelastic material in an arbitrary simple shear flow.  In this first part, we discuss several theoretical aspects of this mathematical formulation and new material property. For example, we show how MAPS measurements can be performed in strain- or stress-controlled contexts and provide relationships between the weakly nonlinear response functions measured in each case. We show that the MAPS response function is a super-set of the response functions that have been previously reported in medium amplitude oscillatory shear and parallel superposition rheology experiments.  We also show how to exploit inherent symmetries of the MAPS response function to reduce it to a minimal domain for straightforward measurement and visualization.  We compute this material property for a few constitutive models to illustrate the potential richness of the data sets generated by MAPS experiments.  Finally, we discuss the MAPS framework in the context of some other nonlinear, time-dependent rheological probes and explain how the MAPS methodology has a distinct advantage over these others because it generates data embedded in a very high dimensional space without driving fluid mechanical instabilities, and is agnostic to the flow protocol.
\end{abstract}


\maketitle




\section{Introduction}


Rheological characterization of materials in the linear viscoelastic regime is well understood. The shear stress and shear strain, for example, can be related in the frequency domain by a complex valued function, $G^*(\omega) = G^\prime(\omega) + iG^{\prime\prime}(\omega)$ \cite{coleman-1961,leaderman-1957}. This function, called the \textit{complex modulus}, can be measured by a variety of strain history protocols, such as SAOS or chirp experiments \cite{tschoegl-1989,geri-2018}. Full knowledge of the complex modulus allows for direct computation of the stress response, $\hat{\sigma}(\omega)$, to an arbitrary strain history protocol, $\hat{\gamma}(\omega)$, through the Boltzmann superposition integral \cite{tschoegl-1989}.

A general approach to nonlinear rheological characterization is not as well developed. Methods exist for representing the nonlinear stress-strain relationship only for specific strain history protocols. Large amplitude oscillatory shear (LAOS), for example, describes this relationship for a single-tone oscillatory strain history \cite{giacomin-1993,mckinley-2008}. Still, the mathematical representation of LAOS can take many forms, such as using Fourier Transform (FT) rheology or using Chebyshev polynomials \cite{hyun-2011,cho-2016}. In either case, the LAOS stress response appears at odd harmonics of the input frequency. Changing the strain history protocol, however, renders LAOS characterization incomplete or ineffective, and other representations are necessary. Parallel superposition (PS) measurements, for example, describe the stress-strain relationship for a strain history protocol consisting of a small amplitude oscillation at a single frequency imposed in the direction of simultaneous steady shear flow \cite{tanner-1968}. Parallel superposition rheology requires an entirely separate representation from LAOS, in which the stress and strain are related by a shear rate-dependent steady shear viscosity and shear rate-dependent complex modulus \cite{vermant-1998,yamamoto-1971}. The specificity of each representation makes direct comparison of LAOS and PS data difficult, and relationships between the measurements are generally only understood in asymptotic limits.

\begin{figure*}[t]
    \centering
    \includegraphics[width = 2\columnwidth]{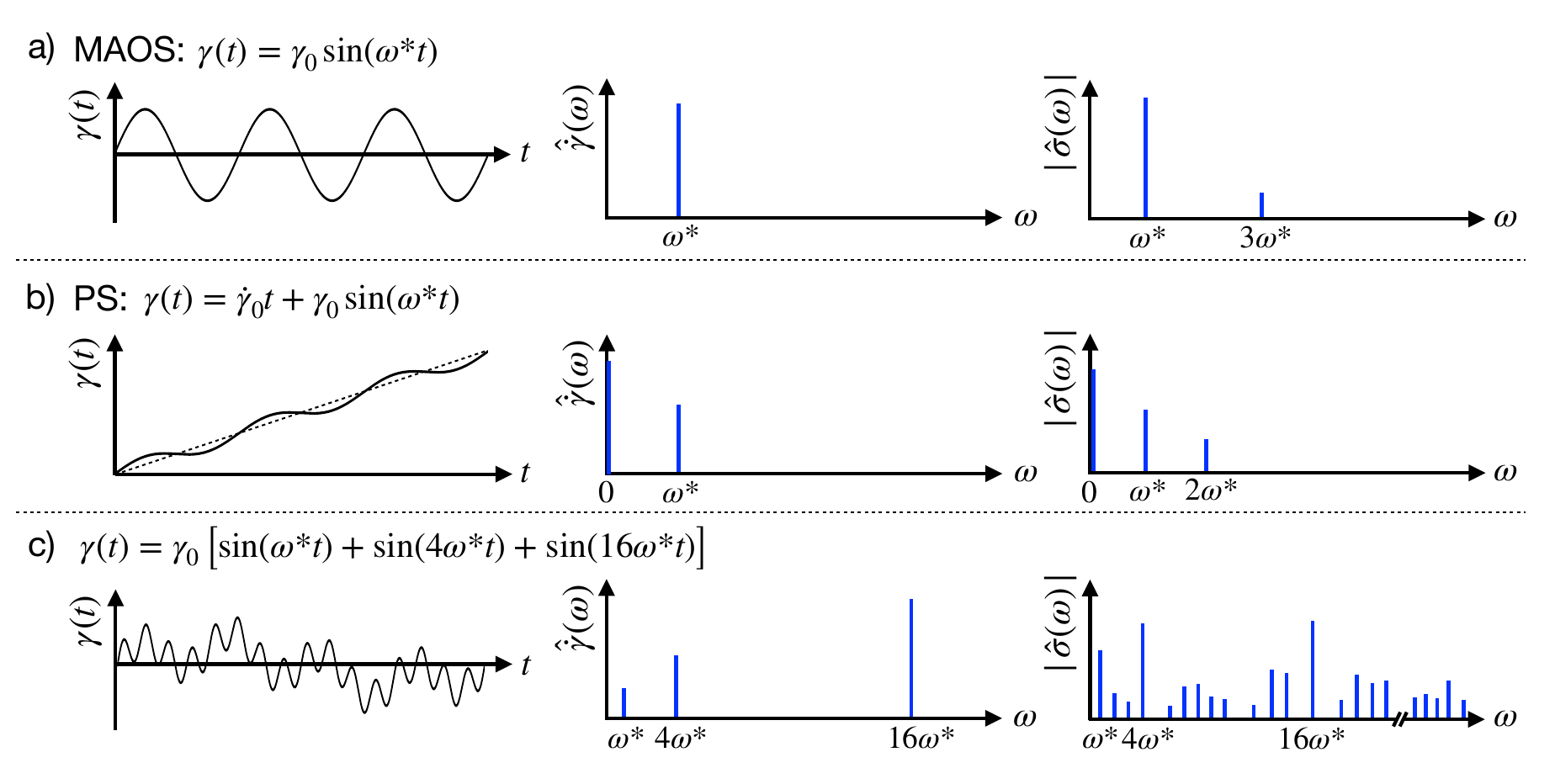}
    \caption{A graphical depiction of three different medium amplitude strain history protocols (left), the Fourier transforms of the corresponding strain rate (center), and the Fourier transforms of the corresponding stress response (right): a) Medium Amplitude Oscillatory Shear (MAOS), b) Parallel Superposition (PS), and c) a waveform consisting of three sine waves imposed in parallel.}
    \label{fig:maos_ps}
\end{figure*}

The issue of generality persists in the weakly nonlinear regime. The medium amplitude analog to LAOS, called medium amplitude oscillatory shear (MAOS), commonly makes use of four intrinsic nonlinear functions: $[e_1](\omega),[v_1](\omega),[e_3](\omega),[v_3](\omega)$, to describe the weakly nonlinear stress response to a single-tone oscillatory strain history \cite{ewoldt-2013}. These functions of the oscillation frequency are sufficient to describe the response at the first and third harmonics of the imposed frequency (Figure \ref{fig:maos_ps}a). Similarly, the existing PS framework is sufficient to describe the steady and first harmonic characteristics of the weakly nonlinear stress response to the superimposed steady-oscillatory flow (Figure \ref{fig:maos_ps}b), using the steady shear viscosity $\eta(\dot{\gamma}_s)$ and shear rate-dependent complex modulus $G^*_{||}(\dot{\gamma}_s,\omega^*)$, respectively. Neither representation, however, can describe the weakly nonlinear stress response to strain history protocols besides the ones for which they are specifically constructed.

Consider, for example, a strain waveform comprised of multiple sine waves imposed in parallel,
\begin{equation}
    \gamma(t) = \gamma_0\sum_{m=1}^{N}\sin{( n_m \omega^*  t + \delta_m)}.
    \label{eq:multitone}
\end{equation}
Compared to MAOS or PS, in the medium amplitude limit the stress response to this imposed strain encodes significantly more information about the underlying relaxation processes in the material. For illustrative purposes, the stress response to a signal with $N = 3$ and a particular set of values for $n_m$, here taken to be $n_1 = 1$, $n_2 = 4$, and $n_3 = 16$, is shown in Figure \ref{fig:maos_ps}c). In particular, the material response will generally possess features not only at the first and third harmonics of the input tones $n_m\omega^*$, but at many intermediate frequencies as well. These additional features arise due to a phenomenon called \textit{intermodulation}, by which a nonlinear system produces outputs at all possible linear combinations of the input tones \cite{volterra-1959}. In the medium amplitude limit, features of the stress response at triplet linear combinations are non-negligible. Thus for the set of input tones $\{\omega^*,4\omega^*,16\omega^*\}$ shown in Figure \ref{fig:maos_ps}c), medium amplitude response features will appear at 22 frequencies: $\omega^*, 2\omega^*, 3\omega^*, 4\omega^*, 6\omega^*, 7\omega^*, 8\omega^*, 9\omega^*, 11\omega^*, 12\omega^*, 13\omega^*, \\ 14\omega^*, 16\omega^*, 18\omega^*, 19\omega^*, 21\omega^*, 24\omega^*, 28\omega^*, 31\omega^*, 33\omega^*, \\ 36\omega^*, 48\omega^*$. Though this example specifies that $N = 3$, intermodulation effects will appear in the medium amplitude response to any signal of the form of equation \ref{eq:multitone} for $N > 1$.

Neither MAOS nor PS, nor any other existing medium amplitude framework, is capable of describing the response to this $N$-tone  waveform. This is true of many other strain history protocols of practical importance, such as square, triangular, or Gaussian waveforms \cite{klein-2007}, all of which can be decomposed as a Fourier sine series of the general form of equation \ref{eq:multitone}. Clearly, the representations of weakly nonlinear rheology that currently exist are not the most general representation possible, indicating the need for a more unified treatment.

In this work, we introduce a representation of nonlinear rheology that is capable of describing the weakly nonlinear response to an arbitrary medium amplitude strain history protocol in terms of only a single function, which we call the \textit{third order complex modulus}, $G^*_3(\omega_1,\omega_2,\omega_3)$. Because the Fourier sine series allows any imposed strain history protocol to be viewed as a series of sine waves applied in parallel, we call this representation Medium Amplitude Parallel Superposition (MAPS) rheology. The third order complex modulus in MAPS is a generalization of the Boltzmann superposition integral to third order in strain (or medium amplitude), which arises from a Volterra series expansion of the general nonlinear functional relating shear stress to shear strain. Both MAOS and PS measurements are simply low-dimensional projections of this general representation for specific strain history protocols (see equations \ref{eq:maos_all} and \ref{eq:ps_all} developed below). 


The mathematical basis underpinning MAPS is a generic polynomial model of nonlinear dynamic responses called a Volterra series \cite{volterra-1959}.  In fields such as electrical and acoustic engineering, the coefficients of Volterra series have proven useful for design of microelectronic \cite{ishikawa-2013,cheng-2017} and acoustic elements \cite{cheng-2017}.  In civil, mechanical and aerospace engineering, Volterra series are used to diagnose failure in mechanical components \cite{chatterjee-2010}.  More broadly, Volterra series act as nonlinear models that can be employed in optimized control schemes \cite{aliyev-2010,cheng-2017}.  In the context of rheology, the Volterra series coefficients represent nonlinear memory functions.  While Volterra series representations of nonlinear viscoelasticity are known from the rational mechanics literature \cite{rivlin-1957}, measurement of an entire nonlinear memory function has never been tackled experimentally \cite{bierwirth-2019}.  However, we will show that MAOS experiments are capable of sampling from a subspace of the manifold on which these functions are defined.  Broadly, the MAPS protocol samples the entire manifold, although how uniformly this space is sampled depends on the specific experimental protocol. These details are discussed in Part 2 of this work. 

In addition to the superior predictive capabilities resulting from the applicability of MAPS to arbitrary strain history protocols, the increased dimensionality of MAPS rheology is desirable from the perspective of nonlinear system identification (NLSI). The usefulness of MAPS for NLSI is enhanced by the advent of new, high throughput experimental techniques, an example of which will also be presented in Part 2 of this work. While a single MAOS experiment, shown in Figure \ref{fig:maos_ps}a, produces only two complex data points (at the input fundamental frequency and the third harmonic), the simple experiment shown in Figure \ref{fig:maos_ps}c allows for the simultaneous measurement of $G^*_3(\omega_1,\omega_2,\omega_3)$ at 19 unique points. These points correspond to 19 of the 22 active channels in the stress response not located at an input first harmonic, which contain only third-order information that can in principle be measured in a single medium-amplitude experiment. More complex shear strain history protocols ($N > 3$)  will further increase the sampled data density of a single MAPS experiment. With access to a variety of data-rich experiments, combined with the proper mathematical formulation to relate and understand them, MAPS is a uniquely suited experimental protocol for application of modern data analysis techniques such as machine learning.


The development of MAPS rheology will be split into two parts: a description of the general theoretical framework and constitutive model examples (Part 1) and analysis and application of detailed experimental protocols (Part 2). Part 1 begins with an introduction of the Volterra series and its application to rheology (section \ref{volterra}), where we develop the Volterra representation for strain, strain rate, and stress-controlled shear protocols. This is followed by a detailed discussion of the medium amplitude regime in particular, in section \ref{MAPS}. This discussion encompasses the mathematical framework of MAPS rheology, in which we demonstrate the relationship between MAPS and existing nonlinear measures, develop a compact method for visualizing MAPS functions and data, and apply the MAPS framework to simple constitutive models. Section \ref{NLSI} then presents a discussion of the implications of MAPS rheology to the critical issue of nonlinear system identification in rheology.

\section{Volterra Series Representation of Weakly Nonlinear Rheology}
\label{volterra}

The shear stress in a viscoelastic material with time-invariant properties undergoing simple shear deformation is a time-invariant functional of the imposed strain (or strain rate) \cite{coleman-1961,rivlin-1959,noll-1958}. When the amplitude of the imposed strain (or strain rate) is sufficiently small, this relationship is linear, and the shear stress, $\sigma(t)$, is represented naturally as the convolution of some material function,  $G(t)$, called the shear relaxation modulus, with the imposed strain history, $\gamma(t)$.  As the strain amplitude is increased, however, the stress-strain relationship becomes nonlinear. To describe this additional complexity, the general nonlinear functional can be expanded as a \textit{Volterra series}, a functional analog to the Taylor series, so that the shear stress is written in the general form:
\begin{align}
\sigma(t) & = \sum_{n=1}^\infty \int \stackrel{n}{\cdots} \int_{-\infty}^t G_n( t - t_1, \ldots, t - t_n ) \label{eq:real} \\
& \quad\quad\quad\quad\quad \times \prod_{m=1}^n \dot \gamma( t_m )  \, d t_m. \nonumber
\end{align}
Here, $ \dot \gamma( t ) $ is the shear rate and the kernels: $ G_n( t - t_1, \ldots, t - t_n ) $, are higher order relaxation moduli analogous to the linear relaxation modulus.  In fact, $ G_1( t ) = G( t ) $.  The nonlinear relaxation moduli have a property analogous to the linear relaxation modulus: causality.  For any $ m = 1,  2, \ldots n $, if $ t - t_m < 0 $, then $ G_n( t-t_1, \ldots, t - t_n ) = 0 $.  It is this property that sets the upper bounds of the integrals in equation \ref{eq:real}.

The Volterra series is simply a generalization of the single convolutional integral expected in linear response to progressively higher orders \cite{volterra-1959,wiener-1942}.  In the context of rheology, it can be thought of as the extension of the Boltzmann superposition principle to higher orders. Just like this well-known principle that governs linear viscoelasticity, the Volterra series is a general representation for all materials with time-invariant properties. The stress response of any material, or any constitutive model, to an arbitrary strain history can be written in the form of equation \ref{eq:real}, though the convergence of the series cannot be guaranteed in all cases, as will be discussed in the following paragraph. Although the Volterra series is known in the rheology community, it has recently been noted \cite{bierwirth-2019} that there have been no attempts to measure any higher order relaxation moduli in full. Notably, equation \ref{eq:real} is a scalar description of the shear stress response, but the stresses in complex fluids are fundamentally tensorial.  The tensorial version of the Volterra series for viscometric flows, called the Frechet series, is known as well \cite{frechet-1910,rivlin-1957, rivlin-1959,bird-1987}.  In the present work, we do not consider normal stresses and so omit this functional relationship for brevity.

The convergence of equation \ref{eq:real} can only be guaranteed for a certain set of input functions. Specifically, a general continuous nonlinear, time-invariant system can be approximated with arbitrary accuracy by a Volterra series if the input functions are restricted to a compact subset of the input function space \cite{franz-2006}. This restriction excludes infinite periodic signals, such as those most commonly used in rheological measurements. However, it has been shown that systems with \textit{fading memory} can be uniformly approximated by a Volterra series even with bounded, infinite-time inputs \cite{boyd-1985} such as sine waves. We leave further discussion of the fading memory criterion in the context of rheology and exploration of those inputs that permit a Volterra series approximation to non-fading memory systems to future works, and consider here only systems that satisfy the fading memory criterion.

\subsection{Strain Controlled Frequency Response\label{sec:strain_control}}

Often measurements of linear viscoelasticity are made in the frequency domain.  The same is true of many popular nonlinear viscoelastic protocols such as MAOS and LAOS.  Therefore, it is useful to write the Volterra series using a frequency space representation.  The Fourier transformation of the shear stress is indicated with a caret:
\begin{equation}
    \hat \sigma( \omega ) = \int_{-\infty}^\infty e^{-i \omega t} \sigma(t) \, dt,
\end{equation}
and likewise for the shear strain and strain rate.  In strain-controlled experiments, the shear stress response of a viscoelastic fluid can be represented as a functional of either the shear strain or the strain rate \cite{noll-1958}. Upon choosing the shear strain, the Volterra series in frequency space is:
\begin{align}
    \hat{\sigma}(\omega) &= \sum_{n = 1}^{\infty} \frac{1}{(2\pi)^{n-1}} \int \stackrel{n}{\cdots} \int_{-\infty}^{\infty}G^{*}_{n}(\omega_1,...,\omega_n) \nonumber \\
    & \quad\quad \quad \quad \times \delta(\omega - \sum_{m=1}^{n}\omega_m) \prod_{m = 1}^{n}\hat{\gamma}(\omega_m) \, d\omega_m
    \label{eq:strain_controlled}
\end{align}
where the kernel is:
\begin{align}
    G^*_n( \omega_1, \ldots, \omega_n ) &= \left( \prod_{m=1}^n i \omega_m \right) \int \stackrel{n}{\cdots} \int_{0}^\infty G_n( t_1, \ldots, t_n ) \nonumber \\
    & \quad\quad\quad\quad\quad \times \prod_{m=1}^n e^{-i \omega_m t_m } \, dt_m.
\end{align}
The reason for choosing the new kernels, $G^{*}_n( \omega_1, \ldots, \omega_n ) $ to represent the Volterra series in frequency space is clear when examining the first term in the series, which is the linear response,
\begin{align}
    \hat{\sigma}(\omega) &= \int_{-\infty}^{\infty}G^{*}_1(\omega_1)\delta(\omega - \omega_1)\hat{\gamma}(\omega_1) \, d\omega_1 + O( \hat \gamma( \omega ) ^2 ) \\
    &= G^{*}_1(\omega)\hat{\gamma}(\omega) + O( \hat \gamma( \omega ) ^2 ). \nonumber
\end{align}
The linear order kernel in this Volterra series, $ G^*_1( \omega ) $ is therefore just the familiar complex modulus from linear viscoelasticity, $G^{*}( \omega ) $. The higher order kernels are generalizations of this function; therefore, we call the $n$th order kernel, $G^{*}_n(\omega_1,...,\omega_n)$, the $n$th order complex modulus.  Its relationship with the $n$th order relaxation modulus $G_n(t_1,...,t_n)$ is just an extension of what is known in linear response.

Equation \ref{eq:strain_controlled} in the context of rheology represents one of the principal developments of this work. To the authors' knowledge, the frequency-domain representation of the Volterra series has not been explored in the field of rheology. In the remainder of this work, this exploration will reveal many interesting properties of the $n$th order complex moduli, and will make clear the relationships between previously disparate experimental frameworks. In Part 2 of this work, further exploration will reveal how the frequency-domain Volterra series permits experiments that can, for the first time, measure the higher order response functions of viscoelastic materials across their entire domain, defined by the $n$-frequency space $(\omega_1,...,\omega_n)$.

Many important features of the $n$th order complex moduli can be directly inferred from symmetries implicit in viscoelasticity and symmetries embedded in the Volterra series itself. Three properties in particular are important for properly understanding and applying the Volterra representation: odd symmetry of the stress with respect to strain, Hermitian symmetry, and permutation symmetry.

\textit{Odd symmetry with respect to strain} --- On changing the sign of the strain: $ \hat \gamma( \omega ) \rightarrow -\hat \gamma( \omega ) $, the sign of the shear stress must also change: $\hat \sigma( \omega ) \rightarrow -\hat \sigma( \omega ) $.  By comparing terms in the Volterra expansions of the original and reversed shear stress, we immediately recognize that:
    \begin{equation}
        G^*_n(\omega_1,...,\omega_n) = (-1)^{n+1} G^*_n(\omega_1,...,\omega_n),
    \end{equation}
    from which we conclude that $ G^*_n(\omega_1,...,\omega_n) = 0 $ when $ n $ is even.  Only odd modes contribute to the Volterra expansion of the shear stress.  This symmetry reduces the number of possible Volterra expansion coefficients by half and establishes the third order coefficient as the leading nonlinear viscoelastic response.
    
    

 \textit{Hermitian symmetry} --- Because $\hat{\sigma}(\omega)$ and $ \hat \gamma( \omega ) $ are the Fourier transformations of real-valued time signals, they are guaranteed to exhibit Hermitian symmetry.  That is, the Fourier transformation of the response at a frequency $-\omega$ is equal to its complex conjugate at the same frequency $\omega$, for example: $ \mathrm{Re}[ \hat{\sigma}(-\omega) ] = \mathrm{Re}[ \hat{\sigma}(\omega) ]$, and $ \mathrm{Im}[ \hat{\sigma}(-\omega) ] = -\mathrm{Im}[ \hat{\sigma}(\omega) ]$.  By comparing terms in the Volterra expansion of the shear stress and employing the mathematical fact that the complex conjugate of a product equals the product of complex conjugates, we conclude that each of the $n$th order complex moduli must also exhibit Hermitian symmetry,
    \begin{align}
        G^{\prime}_n(-\omega_1,...,-\omega_n) = G^{\prime}_n(\omega_1,...,\omega_n),\\
        G^{\prime\prime}_n(-\omega_1,...,-\omega_n) = -G^{\prime\prime}_n(\omega_1,...,\omega_n),
    \end{align}
    where the functions $ G^\prime_n(\omega_1,...,\omega_n) $ and $ G^{\prime\prime}_n(\omega_1,...,\omega_n) $ are the real and imaginary parts of the $n$th order complex moduli.
    
    Hermitian symmetry can be used to understand the $ n $th order complex moduli in the following way.  Changing the sign of the frequency in the response functions is equivalent to reversing the arrow of time.  Elastic responses are invariant to the arrow of time as they depend only on the strain.  However, viscous responses must change sign when the arrow of time reverses, as this also reverses the rate of strain.  Therefore, the Hermitian symmetry indicates that we can unambiguously associate $ G^\prime_n(\omega_1,...,\omega_n) $ with elastic modes of response and $ G^{\prime\prime}_n(\omega_1,...,\omega_n) $ with viscous modes of response.
    
    Additionally, because Hermitian symmetry requires that $ G^{\prime\prime}_n(\omega_1,...,\omega_n) $ be an odd function of frequency, it must vanish when all of its arguments are zero: $ G^{\prime\prime}_n(0,...,0) = 0 $. Because $ G^{\prime}_n(\omega_1,...,\omega_n) $ is an even function, it is not subject to any restriction at zero frequency. This property is consistent with the presence of a delayed elastic effect to a step-strain, but no delayed viscous effect.
   
    Hermitian symmetry of the linear response is commonly exploited in linear viscoelasticity by depicting only the values of $ G^\prime( \omega ) $ and $ G^{\prime\prime}( \omega ) $ at positive values of the frequency when reporting the results of a SAOS frequency sweep.  Due to the Hermitian symmetry of $\hat{\sigma}(\omega)$, the stress response at negative frequencies provides redundant information.  For the $n$th order complex moduli, Hermitian symmetry reduces by half the amount of data needed to describe the entire response function -- in this case by relating the response function at one set of frequencies to the function values at the corresponding negated set of frequencies.  This reduction in data is fortuitous.  However, it is important to remember that there are no generic symmetries relating, for example, $G^*_3(\omega_1,-\omega_2,-\omega_3)$ and $G^*_3(\omega_1,\omega_2,\omega_3)$.  The system response at these two points in frequency space provides unique information about the nonlinear rheology of the material. In contrast with the linear response, consideration of higher order complex moduli at sets of both positive and negative frequencies will be essential.

\textit{Permutation symmetry} --- The final symmetry in the $n$th order complex modulus arises from the mathematical form of the Volterra representation, rather than directly from physical considerations. From Equation \ref{eq:strain_controlled}, the strain dependencies in the Volterra series possess a symmetry with respect to permutation of the particular frequency components.  Therefore, the value of $G^*_n(\omega_1, \ldots, \omega_n ) $ must be invariant with respect to permutation of its arguments.  For example the third order response function satisfies permutation symmetry relations:

\begin{align}
    &G^*_3(\omega_1,\omega_2,\omega_3) = G^*_3(\omega_2,\omega_3,\omega_1) = G^*_3(\omega_3,\omega_1,\omega_2) \nonumber \\
    & \,\, = G^*_3(\omega_3,\omega_2,\omega_1) = G^*_3(\omega_2,\omega_1,\omega_3) = G^*_3(\omega_1,\omega_3,\omega_2).
\end{align}

For the $n$th order complex modulus, permutation symmetry reduces the amount of data required to describe the response function by a factor equal to the number of possible permutations: $ n! $.

Though we have thus far written the Fourier transformation of the shear stress as a Volterra series in the transformed shear strain, it is possible, and sometimes more convenient, to write the stress as a functional of the strain rate history $\dot{\gamma}(t)$. The Volterra series of the transformed stress in this case,
\begin{align}
    \hat{\sigma}(\omega) &= \sum_{n \in \mathrm{odd}} \frac{1}{(2\pi)^{n-1}} \int \stackrel{n}{\cdots} \int_{-\infty}^{\infty}\eta^{*}_{n}(\omega_1,...,\omega_n) \nonumber \\
    & \quad\quad \quad \quad \quad  \times \delta(\omega - \sum_{m=1}^{n}\omega_m) \prod_{m = 1}^{n}\hat{\dot{\gamma}}(\omega_m)d\omega_m, \label{eq:strainrate_controlled}
\end{align}
also depends on odd powers of the strain rate, and the response functions also possess both Hermitian and permutation symmetry. The $n$th order response function $\eta^{*}_n(\omega_1, \ldots, \omega_n) $, which we call the $n$th order complex viscosity, therefore exhibits the same symmetries as does the $n$th order complex modulus. The reason for naming the kernel function in this way is again clear from the familiar relationship between $\eta^*_1(\omega)$ and $G^*_1(\omega)$ in linear viscoelasticity: $ G^*_1(\omega) = i\omega\eta^*_1(\omega) $ \cite{bird-2012}.
A version of this relationship extends to the higher order response functions, which can be inferred by comparing the Volterra expansion with respect to the strain to that with respect to the strain rate:
\begin{equation}
    G_n^{*}(\omega_1,...,\omega_n) = \left( \prod_{m = 1}^{n} i \omega_m \right)  \eta^*_n(\omega_1,...,\omega_n).
    \label{eq:G_to_eta}
\end{equation}
The $n$th order complex viscosity can be written in terms of the $n$th order relaxation modulus:
\begin{align}
 \eta_n^*( \omega_1, \ldots \omega_n ) &=  \int \stackrel{n}{\cdots} \int_{0}^\infty G_n( t_1, \ldots, t_n ) \\ & \quad\quad\quad\quad\quad \times \prod_{m=1}^n e^{-i \omega_m t_m} \, dt_m. \nonumber
\end{align}

Within the multifold integrals of equation \ref{eq:strainrate_controlled}, the Dirac delta function $ \delta( \omega -  \omega_1 - \ldots - \omega_n ) $ has a special influence.  If the $n$th order complex viscosity depends only on the \textit{frequency sum} $:  \omega_1 + \ldots + \omega_n $, and not on the frequencies independently: $ \eta^*_n( \omega_1, \ldots, \omega_n ) \sim \hat f_n( \omega_1 + \ldots + \omega_n ) $, then the relaxation modulus depends only on a single historical time:
\begin{equation}
G_n( t_1, \ldots t_n ) \sim f_n( t_1 ) \prod_{m=2}^n \delta( t_m - t_1 ).
\end{equation}

Finally, because the relaxation moduli are causal, when the decay of $ G_n( t_1, \ldots, t_n ) $ with time $ t_m $, $ m = 1, \ldots n $, is sufficiently fast, the $ n $th order complex moduli are guaranteed to satisfy a set of Kramers-Kronig relations \cite{kronig-1926,kramers-1929,nussenzveig-1987}. For multi-dimensional analytic functions, a number of different relations of this type are known \cite{peiponen-2004,peiponen-2009}.  These results are extensive, and we do not review them here as it is unclear yet whether they serve any useful purpose in the context of rheological measurements.

\subsection{Stress Controlled Frequency Response}

Up to this point, we have restricted the use of the Volterra representation to a strain-controlled formulation. This was done to be consistent with the most common form of rheological experimentation. However, we could equivalently have written the Volterra series in a stress-controlled framework, with the strain or strain rate measured in response to an imposed stress:
\begin{subequations}
\begin{align}
    \hat{\gamma}(\omega) &= \sum_{n \in \mathrm{odds}}  \frac{1}{(2\pi)^{n-1}} \int \stackrel{n}{\cdots} \int_{-\infty}^{\infty}J^{*}_{n}(\omega_1,...,\omega_n) \nonumber \\
    & \quad\quad \quad \quad \quad \times \delta(\omega - \sum_{m=1}^{n}\omega_m) \prod_{m = 1}^{n}\hat{\sigma}(\omega_m)d\omega_m
    \label{eq:stress_controlled_strain} \\
    \hat{\dot{\gamma}}(\omega) &= \sum_{n \in \mathrm{odds}} \frac{1}{(2\pi)^{n-1}} \int \stackrel{n}{\cdots} \int_{-\infty}^{\infty}\phi^{*}_{n}(\omega_1,...,\omega_n) \nonumber \\
    & \quad\quad \quad \quad \quad \times \delta(\omega - \sum_{m=1}^{n}\omega_m) \prod_{m = 1}^{n}\hat{\sigma}(\omega_m)d\omega_m.
    \label{eq:stress_controlled_strain_rate}
    \end{align}
\end{subequations}
In Equation \ref{eq:stress_controlled_strain}, we have chosen the symbol $J^*_n( \omega_1, \ldots, \omega_n ) $ to represent the Volterra kernels because, at first order, $J^*_1( \omega ) $ is the complex compliance from linear response theory. Therefore, we call the kernels, $J^*_n( \omega_1, \ldots, \omega_n )$, the $n$th order complex compliances.  Similarly, in equation \ref{eq:stress_controlled_strain_rate}, the linear response coefficient, $\phi^*_1( \omega )$, is like an inverse complex viscosity. Tschoegl defines this function as the complex fluidity \cite{tschoegl-1989}. Therefore, we refer to $\phi^*_n( \omega_1, \ldots, \omega_n ) $ as the $n$th order complex fluidities.

For experimental studies with gels and yield stress fluids in which it is beneficial to impose the shear stress and measure the resulting shear strain or shear rate, equations \ref{eq:stress_controlled_strain} or \ref{eq:stress_controlled_strain_rate} should be used to relate the stress and deformation. Our analysis of the strain-controlled Volterra kernels, including all symmetries, applies to the stress-controlled kernels as well.

From the point of view of a mildly deformed viscoelastic material, stress and strain control should be indistinguishable. When the stress and strain are both kept at low characteristic amplitudes (i.e. both are kept linear with respect to some amplitude parameter), an oscillatory stress that produces an oscillatory strain can just as well be viewed as an oscillatory strain that produces an oscillatory stress. This is reflected in the well known relationship between $G^*(\omega)$ and $J^*(\omega)$, which extends to our first order complex modulus (viscosity) and compliance (fluidity) functions:
\begin{subequations}
\begin{align}
    J^*_1(\omega) = \frac{1}{G^*_1(\omega)}, \\
    \phi^*_1(\omega) = \frac{1}{\eta^*_1(\omega)}.
\end{align}
\end{subequations}
Similarly, if the stress and strain are both kept as functions that contain a linear component with respect to some characteristic amplitude, and a small cubic component with respect to the amplitude (i.e. what is now commonly referred to as `medium amplitude' forcing), the designation of input versus output also seems arbitrary. Thus, it follows from the first-order inversion relationships that there should be some relationship between the third order functions as well. Indeed, the third order functions are related by
\begin{subequations}
\begin{align}
    & J^*_3(\omega_1,\omega_2,\omega_3) = \label{eq:G_to_J} \\ & \quad -\frac{G^*_3(\omega_1,\omega_2,\omega_3)}{G^*_1(\omega_1)G^*_1(\omega_2)G^*_1(\omega_3)G^*_1(\omega_1+\omega_2+\omega_3)}, \nonumber \\
    & \phi^*_3(\omega_1,\omega_2,\omega_3) = \label{eq:eta_to_phi} \\ & \quad -\frac{\eta^*_3(\omega_1,\omega_2,\omega_3)}{\eta^*_1(\omega_1)\eta^*_1(\omega_2)\eta^*_1(\omega_3)\eta^*_1(\omega_1+\omega_2+\omega_3)}. \nonumber
\end{align}
\end{subequations}
The derivations of these third-order inversion relationships are presented in Appendix \ref{app:stressstrain}.

These inversion relationships are very significant for a few reasons. To the authors' knowledge, no other relationships exist that allow knowledge of the weakly nonlinear response in a strain-controlled experiment to be related directly to the weakly nonlinear response in a stress-controlled experiment. Therefore, these relationships open a conduit between the experimental protocols, and allow data which are collected in very different manners to be directly compared and inter-converted. Additionally, the relationships provide some interesting insight into the third-order nonlinear viscoelastic response. In order to convert -- for example, from a strain-controlled measure ($G^*_3( \omega_1,\omega_2,\omega_3 )$ or $\eta^*_3( \omega_1,\omega_2,\omega_3 )$) to a stress-controlled measure ($J^*_3( \omega_1,\omega_2,\omega_3 )$ or $\phi^*_3( \omega_1,\omega_2,\omega_3 )$) -- then, in addition to the strain-controlled third order function at a given frequency triplet, one must know the linear response at each of the individual frequency coordinates and also at the sum of the frequency coordinates. Therefore, the linear response at the individual frequency components and also at the frequency sum are encoded into the nonlinear response generated at a given point in the general 3D frequency space spanned by $\{\omega_1,\omega_2,\omega_3\}$.

\section{Mathematical Framework for MAPS Rheology}
\label{MAPS}

Up to this point, we have considered all terms in the Volterra series to describe the stress-strain relationship. However, not all terms in the series may be accessible experimentally.  With increasing strain amplitude, many viscoelastic materials spontaneously form shear bands.  In such a case, a relationship like \ref{eq:stress_controlled_strain} is no longer valid because at a given stress level multiple strain histories result.  To avoid such complications, it has recently become common to ensure measurements are made under conditions that are only weakly nonlinear. In this regime, often called the `medium amplitude' regime, it is sufficient to consider only the first and third order terms of the Volterra series. In the remainder of this work, we shall limit our discussion to the first and third order terms only:
\begin{align}
    \hat{\sigma}(\omega) &= G_1^*( \omega ) \hat \gamma( \omega ) \label{eq:strain_controlled3} \\ & + \frac{1}{(2\pi)^2}\iiint_{-\infty}^{\infty}G^{*}_{3}(\omega_1,\omega_2,\omega_3)\delta(\omega - \sum_{j} \omega_j) \nonumber \\
    & \quad \quad \quad \times \hat{\gamma}(\omega_1) \hat{\gamma}(\omega_2) \hat{\gamma}(\omega_3) \, d\omega_1 d\omega_2 d\omega_3 + O( \hat \gamma( \omega )^5)\nonumber.
\end{align}
MAPS  rheology is the process for determining the third order Volterra kernel function $G^*_3(\omega_1,\omega_2,\omega_3)$ from measurements of weak nonlinearities in the shear stress.   

Note, in the sections that follow, we will neglect the $ O( \hat \gamma( \omega )^5 ) $ contributions unless otherwise specified.  Additionally, in the mathematical manipulations to come, summations over the three indices associated with the three dimensional frequency space, $\{\omega_1,\omega_2,\omega_3\}$, will be common.  We will employ the shorthand notations:
\begin{equation}
\sum_j \Longleftrightarrow \sum_{j=1}^3, \quad  \sum_{j \ne k} \Longleftrightarrow \sum_{j=1}^3 \sum_{\substack{k=1\\j\ne k}}^3,  \quad  \sum_{j,k,l} \Longleftrightarrow \sum_{j=1}^3 \sum_{k=1}^3 \sum_{l=1}^3 ,
\end{equation}
to make mathematical expressions more compact.

Table \ref{tab:maps} summarizes the different third order MAPS response functions, their representation as complex numbers, and their associated SI units.  Just as in the linear regime, the dimensionality of the third order response functions indicate how the functions are related. The third order complex modulus and third order complex viscosity differ by a factor of time cubed, indicating that the latter should be multiplied by three factors of frequency to obtain the former, as in equation \ref{eq:G_to_eta}. The third order complex modulus and third order complex compliance differ by dimensions of stress to the fourth power, thus they are related by four factors of the modulus as in equation \ref{eq:G_to_J}. Similar statements can be made about the remaining relationships among the first and third order response functions.  Thus while the $n$th order complex moduli, $ G_1^*( \omega) $ and $G_3^*( \omega_1, \omega_2, \omega_3 ) $ have in common dimensions of stress, the relationships between the other response functions require these quantities to have different dimensions.  Regardless of the simplicity, for some materials or applications it may prove more useful to measure the $n$th order complex viscosity or compliance rather than the modulus itself.

\begin{table*}[t]
    \centering
    \begin{tabular}{c|c|c|c}
        Response function & Complex representation & Definition & SI Units \\
        \hline
        $ G^*_1( \omega ) $ & $ G^\prime( \omega ) + i G^{\prime\prime}( \omega ) $ & Linear complex modulus & Pa \\
        $ \eta^*_1( \omega ) $ & $ \eta^\prime( \omega ) - i \eta^{\prime\prime}( \omega ) $ & Linear complex viscosity & Pa s \\
        $ J^*_1( \omega ) $ & $ J^\prime( \omega ) - i J^{\prime\prime}( \omega ) $ & Linear complex compliance & Pa$^{-1}$ \\
         $ \phi^*_1( \omega ) $ & $ \phi^\prime( \omega ) + i \phi^{\prime\prime}( \omega ) $ & Linear complex fluidity & Pa$^{-1}$ s$^{-1}$  \\
         \hline
         $ G^*_3( \omega_1, \omega_2, \omega_3 ) $ & $ G_3^\prime(  \omega_1, \omega_2, \omega_3 ) + i G_3^{\prime\prime}(  \omega_1, \omega_2, \omega_3 ) $ & Third order complex modulus & Pa \\
        $ \eta^*_3(  \omega_1, \omega_2, \omega_3 ) $ & $ \eta_3^\prime(  \omega_1, \omega_2, \omega_3 ) - i \eta_3^{\prime\prime}(  \omega_1, \omega_2, \omega_3 ) $ & Third order complex viscosity & Pa s$^3$  \\
        $ J^*_3(  \omega_1, \omega_2, \omega_3 )  $ & $ J_3^\prime(  \omega_1, \omega_2, \omega_3 ) - i J_3^{\prime\prime}(  \omega_1, \omega_2, \omega_3 ) $ & Third order complex compliance & Pa$^{-3}$ \\
         $ \phi^*_3(  \omega_1, \omega_2, \omega_3 )  $ & $ \phi_3^\prime(  \omega_1, \omega_2, \omega_3 ) + i \phi_3^{\prime\prime}(  \omega_1, \omega_2, \omega_3 ) $ & Third order complex fluidity & Pa$^{-3}$ s$^{-1}$
    \end{tabular}
    \caption{The measurable response functions for both linear viscoelasticity and the weakly nonlinear material response, representation as complex variables, definitions, and respective SI units.}
    \label{tab:maps}
\end{table*}

Though we have focused primarily on developing the MAPS framework for strain-controlled experiments, employing the third order complex modulus $G^*_3(\omega_1,\omega_2,\omega_3)$, it is sometimes more convenient to work in the strain rate-controlled framework, with the third order complex viscosity $\eta^*_3(\omega_1,\omega_2,\omega_3)$. In the forthcoming discussion of constitutive models, this is often the case. The models explored in this work are predominantly viscous at low oscillation frequencies, thus the complex viscosity is the more natural descriptor. In the remainder of this work, we present expressions in terms of either material function, based on which function arises more naturally in the situation at hand. For any expression that is written solely in terms of $\eta^*_3(\omega_1,\omega_2,\omega_3)$, equation \ref{eq:G_to_eta} makes it easy to find the corresponding expression in terms of $G^*_3(\omega_1,\omega_2,\omega_3)$ if desired, and vice versa.

\subsection{Relationship to Medium Amplitude Oscillatory Shear (MAOS)}
\label{sec:maos}

To begin understanding how the functions $G^*_3( \omega_1, \omega_2, \omega_3 ) $ and $\eta^*_3( \omega_1, \omega_2, \omega_3 ) $ should be interpreted physically, it is instructive to consider the general weakly nonlinear stress response of a viscoelastic system to some simple strain histories. One canonical signal is a \textit{single-tone} oscillatory strain, $\gamma(t) = \gamma_0\sin(\omega t)$. The weakly nonlinear response to this oscillatory strain is most clearly described by the four MAOS material functions proposed by Ewoldt and Bharadwaj \cite{ewoldt-2013},
\begin{align}
    \sigma(t) &= \gamma_0 \left[ G^\prime( \omega ) \sin( \omega t ) + G^{\prime\prime}( \omega ) \cos( \omega t ) \right] \\ &+ \gamma_0^3\Big([e_1](\omega)\sin(\omega t) + \omega[v_1](\omega)\cos(\omega t) \nonumber \\ 
    & - [e_3](\omega)\sin(3\omega t) + \omega[v_3](\omega)\cos(3\omega t)\Big), \nonumber
\end{align}
though similar representations have been previously proposed \cite{davis-1978}.

For this simple oscillatory strain history, the third-order term in the Volterra series can be integrated analytically, and the response examined in Fourier space at different harmonics of the driving oscillation frequency. To third order in the strain amplitude, the single-tone strain excites a stress response at the first and third harmonics only.
By comparing the expression for the shear stress obtained from the MAPS framework, via substitution of the Fourier transform of the input signal into equation \ref{eq:strain_controlled}, with the Fourier transform of the above expression from the MAOS framework, each of the four MAOS material functions can be directly related to either $G^*_3(\omega_1,\omega_2,\omega_3)$ or $\eta^*_3(\omega_1,\omega_2,\omega_3)$:
\begin{subequations}
\label{eq:maos_all}
    \begin{align}
    [e_1](\omega) &= \frac{3}{4} G^\prime_3(\omega,-\omega,\omega) = \frac{3\omega^3}{4} \eta^{\prime\prime}_3(\omega,-\omega,\omega),
    \label{eq:e1}\\
    [v_1](\omega) &= \frac{3}{4\omega}G^{\prime\prime}_3(\omega,-\omega,\omega) = \frac{3\omega^2}{4} \eta^{\prime}_3(\omega,-\omega,\omega),
    \label{eq:v1}\\
    [e_3](\omega) &= \frac{1}{4} G^{\prime}_3(\omega,\omega,\omega) = -\frac{\omega^3}{4} \eta^{\prime\prime}_3(\omega,\omega,\omega),
    \label{eq:e3}\\
    [v_3](\omega) &= -\frac{1}{4\omega}G^{\prime\prime}_3(\omega,\omega,\omega) = \frac{\omega^2}{4} \eta^{\prime}_3(\omega,\omega,\omega).
    \label{eq:v3}
    \end{align}
\end{subequations}
By analogy with the physical interpretation of the MAOS material functions, equations \ref{eq:e1} through \ref{eq:v3} demonstrate that, for the single tone oscillatory input, the real component of the three dimensional complex function $G^*_3( \omega_1, \omega_2, \omega_3 )$ (or the imaginary component of $\eta^*_3( \omega_1, \omega_2, \omega_3 )$) sampled at a special choice of $\{\omega_1,\omega_2,\omega_3\}$ corresponds to the elastic material response in MAOS, and the imaginary component of $G^*_3( \omega_1, \omega_2, \omega_3 )$ (or equivalently the real component of $\eta^*_3( \omega_1, \omega_2, \omega_3 )$) corresponds to the viscous material response measured in MAOS.

Equations \ref{eq:e1} through \ref{eq:v3} also demonstrate the more general nature of the Volterra representation over the MAOS material function representation. The four distinct MAOS material functions are clearly different projections of the same general underlying nonlinear response, which is not immediately apparent from the MAOS representation alone. The relationship between the third and first harmonic responses, and between the elastic and viscous component of the response at either harmonic, appears straightforwardly in the Volterra representation, however. The Volterra representation also indicates that the single tone MAOS response is a subset of a much broader class of weakly nonlinear material responses.  That is, the single tone response is just a special case where $|\omega_1| = |\omega_2| = |\omega_3| = \omega $. Thus, while the MAOS material functions cannot be easily interconverted to provide the material response for strain inputs other than single tone oscillatory deformation histories, the generality of the Volterra expansion makes such relations simple.  The expressions in equations \ref{eq:e1}-\ref{eq:v3} can be used to derive the previously unknown inverse relationships between the intrinsic nonlinear functions in stress-controlled and strain-controlled MAOS by combining them with equations \ref{eq:G_to_J} and \ref{eq:eta_to_phi}. The resulting expression are difficult to express compactly; therefore we report them in Appendix \ref{app:maostress}.

\subsection{Relationship to Parallel Superposition (PS)}
\label{sec:ps}

A slightly more complex shear strain history that is used to probe nonlinear viscoelasticity consists of a superimposed steady shear and single tone oscillation. When the oscillation acts in the same direction as the steady shear flow, the experiment is termed parallel superposition (PS) \cite{dealy-1999}, and the strain can be written in the form:
\begin{equation}
    \gamma(t) = \dot{\gamma}_{s}t + \gamma_0\sin(\omega t),
\end{equation}
where $\dot{\gamma}_s$ represents the steady shear rate, and $\gamma_0$ represents the amplitude of the oscillations. These parameters can be varied separately, with $\gamma_0$ usually being kept small while $\dot{\gamma}_s$ is varied to excite nonlinearities in the material response (as shown in Figure \ref{fig:maos_ps}). Using the Volterra series representation of equation \ref{eq:strainrate_controlled}, in the limit of small $ \dot \gamma_s $ and $ \gamma_0 \omega $, the stress to third order can be written as:
\begin{align}
    &\sigma( t ) =  \dot \gamma_s \left( \eta_1^*( 0 ) + \dot \gamma_s ^ 2 \eta_3^*( 0, 0, 0 ) + \frac{3}{2} \gamma_0^2 \omega^2 \eta_3^*( \omega, -\omega,0 )  \right) \nonumber \\
    &\quad + \gamma_0 \omega \left( \eta^{\prime\prime}_1( \omega ) + 3 \dot \gamma_s^2 \eta^{\prime\prime}_3( \omega, 0, 0 ) \right) \sin (\omega t) \nonumber \\
    &\quad + \gamma_0 \omega \left( \eta^{\prime}_1( \omega ) + 3 \dot \gamma_s^2 \eta^{\prime}_3( \omega, 0, 0 ) \right) \cos(\omega t) \nonumber \\
    &\quad + \frac{3}{2} \dot \gamma_s \gamma_0^2 \omega^2 \left( \eta^{\prime\prime}_3( \omega, 0, \omega ) \sin (2 \omega t) + \eta^{\prime}_3( \omega, 0, \omega ) \cos (2 \omega t) \right)  \label{eq:ps}
\end{align}
By convention in parallel superposition rheology, one identifies an apparent steady shear viscosity, $ \eta_s( \dot \gamma_s, \gamma_0, \omega ) $ and components of the stress in and out of phase with the imposed oscillation, which when normalized by $ \gamma_0 $ are termed the parallel complex modulus, $ G^*_\parallel( \dot \gamma_s, \gamma_0, \omega ) $ \cite{yamamoto-1971}.  On examination of equation \ref{eq:ps}, we find that these two quantities can be related to the first and third order complex viscosities contained in the Volterra series by the following equalities, valid for weakly nonlinear PS flows:
\begin{subequations}
\begin{align}
    \eta_s( \dot \gamma_s, \gamma_0, \omega ) &= \eta^*_1(0) + \dot{\gamma}_s^2\eta^*_3(0,0,0) \label{eq:ps_eta}  \\
    & \quad + \frac{3}{2} \gamma_0^2 \omega^2 \eta_3^*( \omega,-\omega,0 ),\nonumber
    \\
    G^*_{||}( \dot \gamma_s, \gamma_0, \omega ) &= i\omega\left(\eta^*_1(\omega) + 3\dot{\gamma}_s^2\eta^*_3(\omega,0,0)\right).
    \label{eq:ps_G}
\end{align}
    \label{eq:ps_all}
\end{subequations}
It is worth noting that the linear combinations of first and third order complex moduli measured in PS measurements are distinct from those measured in MAOS.  Although the steady shear viscosity (equation \ref{eq:ps_eta}) is written in terms of the first and third order complex viscosities, the Hermitian and permutation properties of the $n$th order complex viscosities guarantee that the imaginary components all vanish; therefore, the steady shear viscosity is indeed a real-valued quantity.  The stress response in equation \ref{eq:ps} also contains a term that is a second harmonic of the oscillation, but is not typically measured experimentally.  Of note is the fact that the viscosities, $ \eta_3^*( 0,0,0) $, $ \eta_3^*(\omega,-\omega,0 )$, and $\eta_3^*( \omega,0,\omega )$ are all associated with even order harmonics of the oscillation frequency in equation \ref{eq:ps}, and their arguments depend on the driving frequency an even number of times.  In contrast, $ \eta_3^*( \omega, 0, 0 ) $ is associated with an odd harmonic in the PS experiment, and its argument depends on the driving frequency an odd number of times.  This difference in even-odd dependence relates in a significant way to the geometry of the third order response functions and is discussed in more detail in section \ref{sec:geometry}. 

Equations \ref{eq:ps_eta} through \ref{eq:ps_G} illustrate the power of the general Volterra representation in concisely capturing more complex behavior than simple single tone oscillations. The same underlying material functions can indeed represent the pure oscillatory response and the superimposed steady shear-oscillatory shear response. Equations \ref{eq:ps_eta} and \ref{eq:ps_G} also present a simple interpretation for some phenomena associated with parallel superposition. Experiments have shown that the parallel complex modulus can become negative for low oscillation frequencies \cite{yamamoto-1971}. From Equation \ref{eq:ps_G}, we see that this sign change is associated with an increased dominance in the third-order nonlinearity when it has opposite sign to the linear response at these frequencies. It has also been noted that common relations between the steady shear stress and the zero frequency modulus fail for PS measurements \cite{vermant-1998}
\begin{equation}
    \lim_{\omega\xrightarrow{}0}\left(\frac{\text{Im}\left[G^*_{||}( \dot \gamma_s, \gamma_0, \omega )\right]}{\omega}\right) \neq \lim_{\omega\xrightarrow{}0}\eta_s( \dot \gamma_s, \gamma_0, \omega ).
\end{equation}
This clearly is true from Equations \ref{eq:ps_eta} and \ref{eq:ps_G}; the latter expression possesses a factor of 3  multiplying the third order complex viscosity,  $\eta_3^*(0,0,0) $, while the former has only a factor of unity multiplying the same.  This difference ultimately arises from permutation symmetry.

\begin{figure*}[t]
    \centering
    \includegraphics[width=0.75\textwidth]{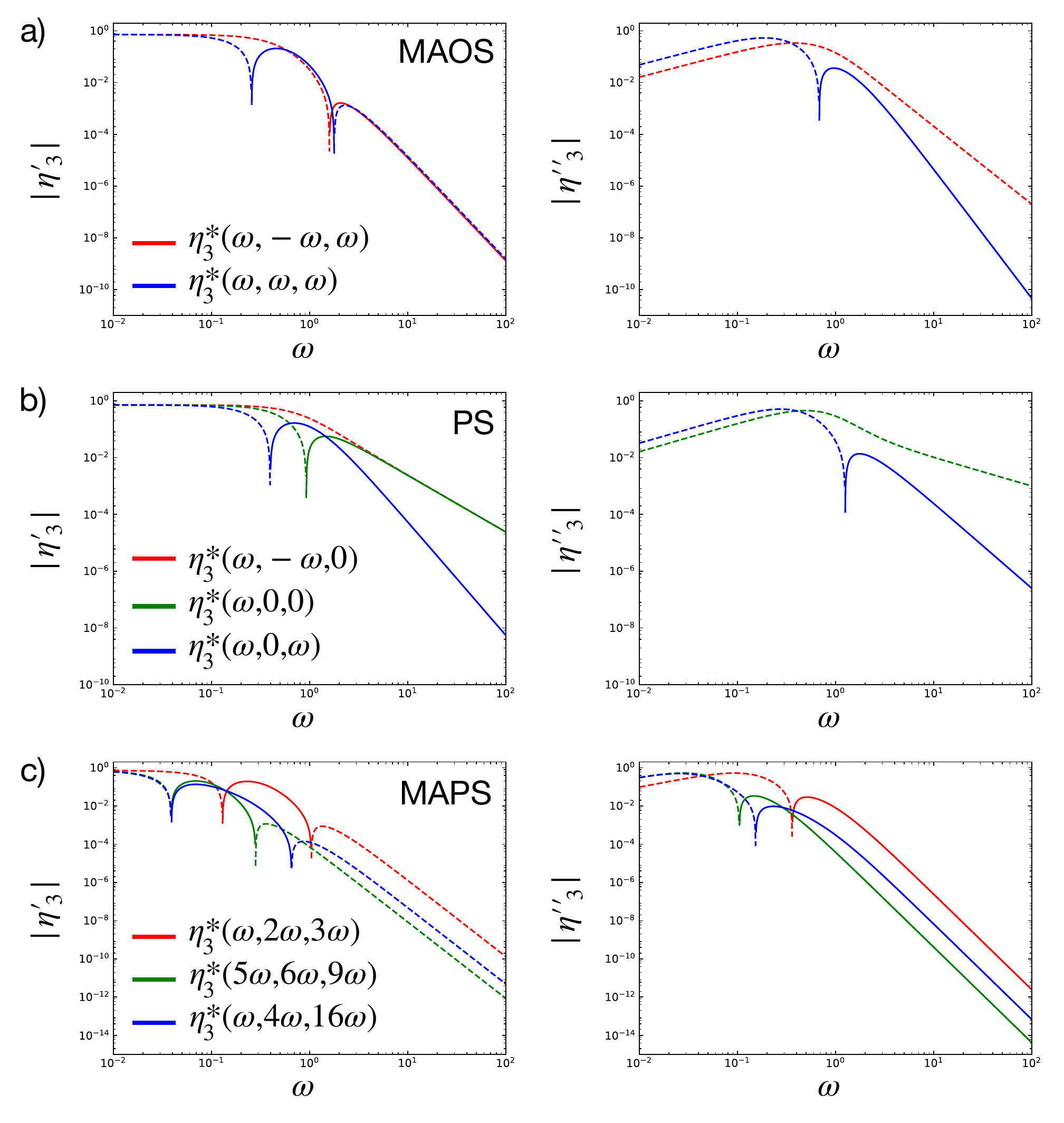}
    \caption{Projections of the third order complex viscosity of a Giesekus fluid with $\alpha = 0.3$ and all remaining phenomenological constants set to unity. The left-hand column shows the real (viscous) response and the right-hand column shows the imaginary (elastic) response. Solid lines represent positive values of the functions, and dashed lines represent negative values.  (a) Projection of the 3D complex MAPS function $\eta^*_3(\omega_1,\omega_2,\omega_3)$ as measured by MAOS, which samples the function at $\omega_1 = \omega_2 = \omega_3 = \omega$ and $\omega_1 = -\omega_2 = \omega_3 = \omega$ (see equations \ref{eq:e1}-\ref{eq:v3}). (b) Projection of $\eta^*_3(\omega_1,\omega_2,\omega_3)$ as measured by PS, which samples the function at $\omega_1 = -\omega_2 = \omega$, $\omega_3 = 0$; $\omega_1 = \omega$, $\omega_2 = \omega_3 = 0$; and $\omega_1 = \omega_3 = \omega$, $\omega_2 = 0$ (see equations \ref{eq:ps_eta}-\ref{eq:ps_G}). (c) Projection of $\eta^*_3(\omega_1,\omega_2,\omega_3)$ at other coordinates within the 3D MAPS domain for which $\{\omega_1,\omega_2,\omega_3\} = \{n_1, n_2, n_3\} \times \omega$, with $\{n_1,n_1,n_3\} = \{1,2,3\}$, $\{5,6,9\}$, and $\{1,4,16\}$, respectively. Note that the apparent singularities in the above curves represent a change in sign of the real or imaginary component of $\eta^*_3(\omega_1,\omega_2,\omega_3)$.}
    \label{fig:giesekus}
\end{figure*}

\subsection{Visualization of MAPS Response Functions}
\label{sec:geometry}

For linear viscoelasticity, multiple methods of visualizing the complex modulus or complex viscosity exist. For example, the magnitude and phase angle of the complex modulus might be plotted against the frequency as in a Bode plot. Or, the real and imaginary parts of the modulus might be plotted against one another as in a Nyquist (Cole-Cole) plot. MAOS and PS data, whose domain is a single frequency dimension, are amenable to the same visualization strategies. Such two dimensional visualizations are a crucial tool that enables scientists and engineers to learn from data collected via experiment, simulation, or theory.

As we discussed in detail in sections \ref{sec:maos} and \ref{sec:ps}, MAOS and PS measure specific projections of the more general MAPS response functions. Within the three-dimensional domain: $\{\omega_1,\omega_2,\omega_3\}$, these common protocols measure points at which $\{\omega_1,\omega_2,\omega_3\} = \{n_1, n_2, n_3\} \times \omega$ for fixed integer triplets $\{n_1, n_2, n_3\}$ while sweeping the fundamental frequency $ \omega $. A MAOS experiment probes $\{n_1,n_2,n_3\} = \{1,1,1\}$ and $\{1,-1,1\}$, and a PS experiment probes $\{n_1,n_2,n_3\} = \{1,-1,0\}$, $\{1,0,0\}$, and $\{1,0,1\}$. In other words, from the three-dimensional domain of MAPS, these protocols sample along different one-dimensional manifolds -- lines -- spanned by the single frequency coordinate: $\omega$, and  distinguished by the values of the triplet: $\{n_1,n_2,n_3\}$. Figures \ref{fig:giesekus}(a) and (b) provide an example visualization of the MAOS and PS responses for the Giesekus constitutive model, which have been obtained from the solution for $\eta^*_3(\omega_1,\omega_2,\omega_3)$ presented later in section \ref{sec:giesekus}. The curves depicted in Figure \ref{fig:giesekus}(a) and (b) agree with previously obtained solutions for the Giesekus model in MAOS \cite{gurnon-2012} and PS \cite{kim-2013}.

It is clear from the previous discussion that MAOS and PS are not unique or privileged as descriptors of the nonlinear response of a complex fluid. It is possible, for example, to select other one-dimensional projections of a MAPS response function by simply choosing $\{n_1,n_2,n_3\}$ to take values other than those taken in MAOS or PS. These projections are still functions of a single frequency variable $\omega$ and can be visualized in the same manner as MAOS or PS. Figure \ref{fig:giesekus}(c) demonstrates that, for a Giesekus fluid, the projections defined by $\{n_1,n_2,n_3\} = \{1,2,3\}$, $\{5,6,9\}$, and $\{1,4,16\}$ have a dependence on $\omega$ that is distinctive from the projections that appear in MAOS or PS. The rationale for choosing these particular frequency combinations will become clear in Part 2 of this work.

The visualization strategy employed in Figure \ref{fig:giesekus} is beneficial in that it is familiar from visualization of linear response data, and that it allows the complex MAPS response functions to be depicted as functions of a single common experimental variable, $\omega$. However, this strategy is insufficient to capture the richness of high-dimensional MAPS data. Figure \ref{fig:giesekus} says nothing of the relationship between the different 1D manifolds parameterized by $\{n_1,n_2,n_3\}$. Relating these manifolds that are accessed by different experimental protocols, such as MAOS and PS, is a principal development of the MAPS framework, which any visualization strategy should reflect. Moreover, should we want to observe the functional dependence of the MAPS response function on the coordinates $\{n_1,n_2,n_3\}$ rather than on $\omega$, the strategy of Figure \ref{fig:giesekus} is ineffective. To accomplish these objectives, a new, more complete visualization strategy is necessary. The remainder of this section is devoted to developing such a strategy for MAPS rheology.


To truly quantify the weak nonlinearities in an unknown material we must probe the three dimensional frequency space of the MAPS response function broadly.  This, however, presents some challenges with regard to visualization and interpretation of high dimensional data.  At third order, the MAPS response function is a map from the three dimensional frequency space, $ \mathbb{R}^3 $ to the space of complex numbers, $ \mathbb{C} $ -- a volume embedded in a five dimensional space.  Ideally, we would like a method to visualize the entirety of this set of data in the two dimensional plane.  To enable this kind of visualization, we need to reduce the manifold on which the MAPS response functions are measured to a minimal set of subspaces by utilizing the symmetries discussed in section \ref{sec:strain_control}.


    
    

\begin{figure*}[t]
    \centering
    \includegraphics[width=1.8\columnwidth]{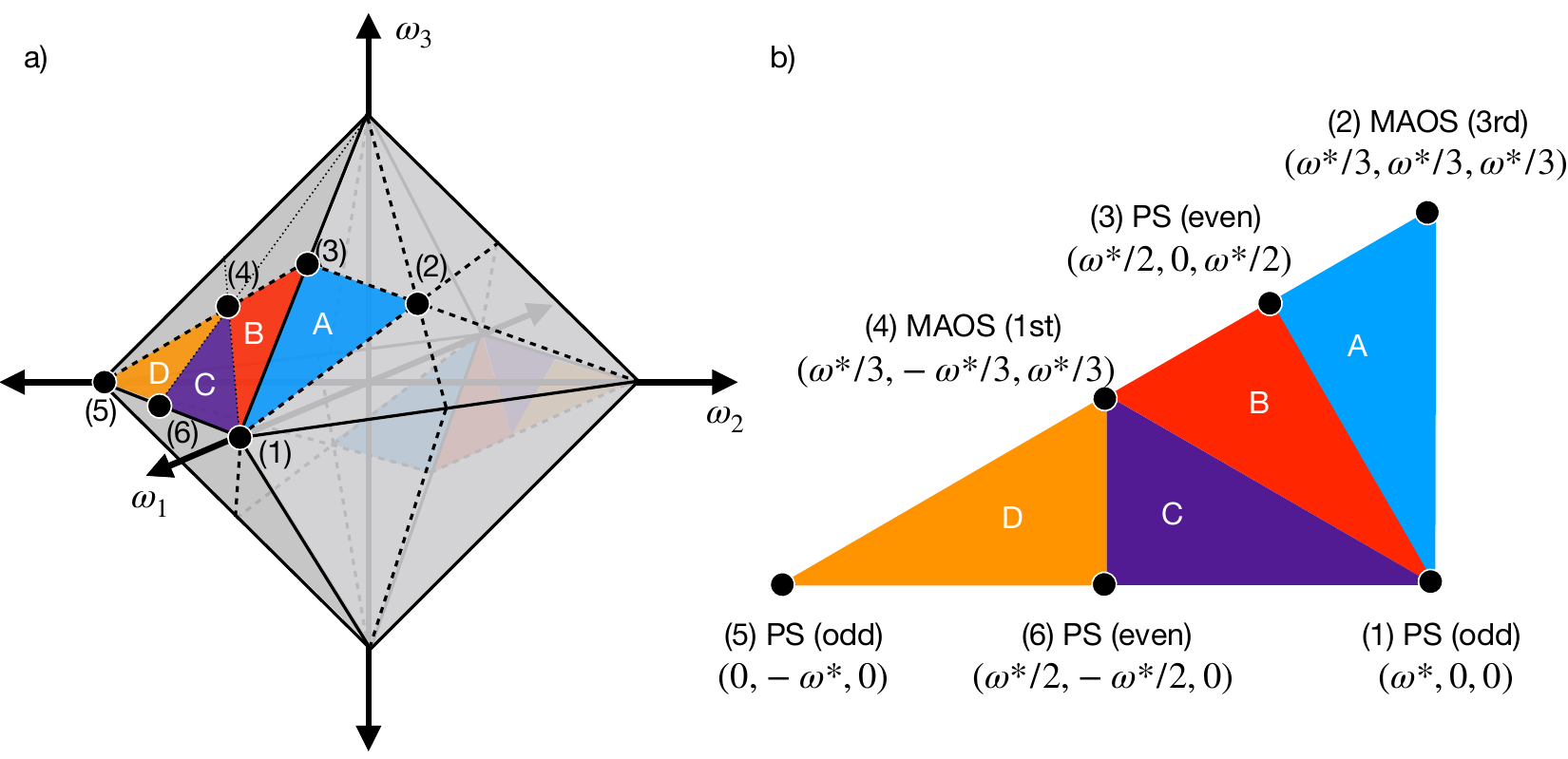}
    \caption{a) A constant $L^1$-norm manifold in 3D frequency space is octahedral, $ \| \boldsymbol{\omega} \|_1 = \omega^* $.  The surface of the octahedron has four triangular subspaces: A, B, C, D, within which the third order complex modulus can be identified uniquely.  Dotted lines indicate planes associated with permutation symmetry.  Reflection of the triangles by Hermitian symmetry is indicated on the back of the octahedron.  b) The unique triangular subspaces can be projected as a similar right triangle on the plane.  Each vertex of a triangular subspace corresponds to a point measurable by either MAOS or PS rheology.}
    \label{fig:octahedron}
\end{figure*}

Identification of these subspaces is made easiest by examining a constant $L^1$-norm surface in the three dimensional frequency space formed by the vector $\boldsymbol{\omega} = (\omega_1, \omega_2, \omega_3)$,
\begin{equation}
    \| \boldsymbol{\omega} \|_1 = | \omega_1 | + | \omega_2 | + | \omega_3 | = \omega^*.
\end{equation}
This surface is just a regular octahedron, and is depicted in Figure \ref{fig:octahedron}a. Permutation symmetries indicated by the dashed lines divide the octahedral surface into 12 subregions of which only two can be specified uniquely.  Two unique regions are shaded on the front and back of the octahedron.  However, even these two shaded regions are related directly by the Hermitian symmetry of the response function.  Therefore, the behavior of the third order complex modulus on a constant $L^1$-norm surface can be captured entirely by the values it takes in the front facing shaded region in Figure \ref{fig:octahedron}a).

In Figure \ref{fig:octahedron}b), this shaded region is lifted off the surface of the octahedron and laid flat in the plane.  The shaded region can be further decomposed into four right-triangular subspaces.  We label these subspaces, $ A $, $ B $, $ C $, and $ D $ to distinguish them. The regions are geometrically defined by the inequalities:

\begin{subequations}
\begin{align}
    &\text{A:} \quad \omega_1 \geq \omega_3 \geq \omega_2, \quad \omega_1,\omega_2,\omega_3 \geq 0, \\
    &\text{B:} \quad \omega_1 \geq \omega_3 \geq -\omega_2, \quad \omega_1,\omega_3 \geq 0 \geq \omega_2, \\
    &\text{C:} \quad \omega_1 \geq -\omega_2 \geq \omega_3, \quad \omega_1,\omega_3 \geq 0 \geq \omega_2, \\
    &\text{D:} \quad -\omega_2 \geq \omega_1 \geq \omega_3, \quad \omega_1,\omega_3 \geq 0 \geq \omega_2.
\end{align}
\label{eq:subspaces}
\end{subequations}

The choice of $\omega_2 \leq 0$ in subspaces $B$, $C$, and $D$ in Figure \ref{fig:octahedron} is arbitrary; four identically-shaped subspaces could have been drawn elsewhere on the surface of the octahedron to capture the same behavior as those described by equation \ref{eq:subspaces}. Proper application of permutation and Hermitian symmetries, however, allows any point in 3D frequency space to be associated with a point that satisfies one of the four inequalities in equation $\ref{eq:subspaces}$.

The subspaces are identical geometrically (they are all hemi-equilateral triangles), but the values of the third order complex modulus $G^*_3(\omega_1,\omega_2,\omega_3)$ within each triangle can take on different values.  Each triangular subspace has three vertices associated with the weakly nonlinear viscoelastic tests discussed in sections \ref{sec:maos} and \ref{sec:ps}.  Each subspace will have one vertex that is measured via MAOS, one measured by the odd harmonic term in PS (equation \ref{eq:ps}), and one measured by an even harmonic in PS.  The MAOS vertex always resides at the sixty degree angle, and the even harmonic PS vertex always resides at the ninety degree angle.  From this geometric decomposition of the space, it is apparent that the MAOS and PS responses that rheologists measured and reported describe nonlinearity only along the periphery of the unique subspaces for the more general third order response functions.  As far as the authors are aware, there have been no rheological investigations of the interior of this domain.  As we shall demonstrate, the periphery provides a good description of the entire domain only at sufficiently low frequencies, $\omega^*$.

The benefits of this geometric representation are clear. The constant $L^1$-norm manifold has been broken down into just four distinct regions, each of which occupy only $1/48$th of the entire manifold.  Measurements of the third order response function made at other positions on the octahedral surface can be mapped back into just these four subspaces using the appropriate permutation and Hermitian symmetries.  This is a massive reduction in complexity.  The simplest visualizations of the response function are contour plots of the real and imaginary parts of third order response function within the bounding triangle shown in Figure \ref{fig:octahedron}b).  If the response function is continuous, the contours are guaranteed to be continuous within this region and can be evaluated repeatedly for different values of the frequency $ L^1 $-norm.  For example, later in section \ref{sec:examples}, Figures \ref{fig:low_freq} through \ref{fig:giesekus_MAPS} give examples of these contours.  This visualization can be realized when an analytical expression or numerical solution for the third order response function is known.  Experimentally it may prove difficult to collect data densely enough to populate the constant $ L^1 $-norm surface and then generate the corresponding contour plot.  In such a case, an alternative visualization scheme is possible, in which MAPS data is presented using the familiar Bode and Nyquist plots often used to display SAOS and MAOS data. The details of this scheme are specified below.

\begin{figure}[t]
    \centering
    \includegraphics[width=\columnwidth]{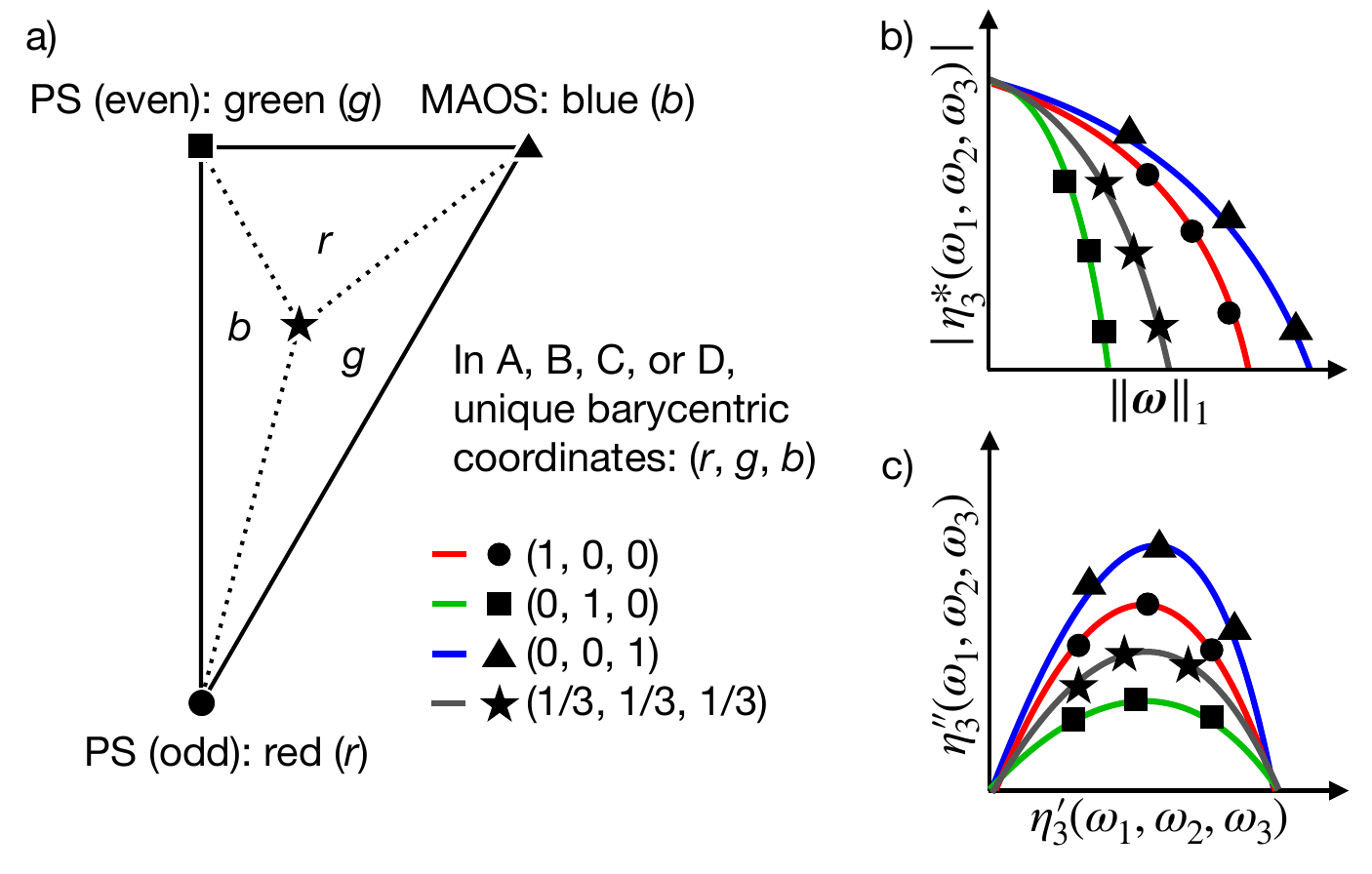}
    \caption{Depiction of a barycentric visualization scheme for each triangular subspace. (a) Within one of the four unique triangular subspaces a set of barycentric coordinates $(r,g,b)$ can be constructed.  The position of data within this triangle can be associated with a color using the $ (r,g,b) $ coordinates as color channels (or with a unique symbol). b) A Bode plot of the third order complex viscosity as a function of the $L^1$ norm of the frequency.  Each curve or symbol is associated with a different point within the triangular subspace as indicated in a). c) A Nyquist plot of the real and imaginary parts of the third order complex viscosity $\eta^*_3(\omega_1,\omega_2,\omega_3)$ measured at different values of $ \| \omega \|_1 $.  As with b) different curves and symbols are associated with different points with the triangular subspace.}
    \label{fig:barycentric}
\end{figure}

Because these subspaces are identically shaped, they are amenable to visualization schemes that treat the subspaces individually. Figure \ref{fig:barycentric}a) depicts one of the four triangular subspaces.  Which subspace is irrelevant because they are all identical geometrically.  The different flows corresponding to each vertex in this hemi-equilateral triangle are also identified.  A point within the subspace can be identified uniquely by its barycentric coordinate: $(r,g,b)$, where the numbers $ r, b, g \in [0,1] $ with $ r + g + b = 1 $, describe how close the point is to one of the three distinct vertices.  The barycentric coordinate system is constructed simply by computing the relative area of the triangular subdivisions formed when the point of interest is connected to each of the vertices.  We establish the convention that:
\begin{itemize}
    \item the coordinate with $ r = 1 $ specifies the odd PS ($30^{\circ}$) vertex, 
    \item the coordinate with $ g = 1 $ specifies the even PS ($90^{\circ}$) vertex,
    \item the coordinate with $ b = 1 $ specifies the MAOS ($60^{\circ}$) vertex.
\end{itemize}
Thus, a barycentric coordinate $(0,9, 0.05, 0.05)$ is close to the MAOS vertex and a point $(0.\overline{3}, 0.\overline{3}, 0.\overline{3})$ is the barycenter of the triangle.

Visualization can be performed for each of the triangular subspaces $ A $, $ B $, $ C $, or $ D $ individually by creating Bode or Nyquist plots (as shown in Figures \ref{fig:barycentric}(b) or \ref{fig:barycentric}(c) respectively) of the response function at particular values of the barycentric coordinate while steadily sweeping the frequency $L^1$-norm, $ \| \boldsymbol{\omega} \|_1 $.  Particular symbols or line colors can be associated with each different barycentric coordinate.  A particular advantage of representing the data in terms of the coordinates $(r,g,b)$ is that the values of coordinates can be used as color channels in an RGB color scheme to naturally and uniquely color the data in the plot.  In such a case, red lines are close to odd PS vertices, green lines are close to even PS vertices, and blue lines are close to MAOS vertices, regardless of the specific triangular subspace being analyzed.  This gives a common visual language to the general third order response function in each triangular subspace. In Figures \ref{fig:fluidity} through \ref{fig:giesekus_MAPS}, examples of MAPS sweeps at select barycentric coordinates for various constitutive models are presented, which have been colored using the scheme outlined above.

\subsection{Computing the MAPS Response Function for Some Phenomenological and Constitutive Models\label{sec:examples}}

There is a richness in the range of different MAPS responses possible for viscoelastic materials.  In this work, we will not investigate in depth the responses arising from different classes of constitutive models.  Instead, we look broadly at six phenomenological examples for which the MAPS response is relatively simple to compute and understand without regard for micro-mechanical details.

\subsubsection{Phenomenological Model: Low Frequency Response of Viscoelastic Fluids}
\label{sec:lowfreq}

In the limit of low frequencies, viscoelastic fluid materials with a longest relaxation time exhibit a very simple third order complex viscosity:
\begin{equation}
    \eta_3^*( \omega_1, \omega_2, \omega_3 ) = a + i b \sum_{j} \omega_j + \frac{c}{2} \sum_{j \ne k} \omega_j \omega_k + d \sum_j \omega_j ^ 2, \label{eq:lowfreq}
\end{equation}
which is characterized by four independent coefficients that we denote: $ a, b, c, d $ (with SI units of Pa$\cdot$s\textsuperscript{3}, Pa$\cdot$s\textsuperscript{4}, Pa$\cdot$s\textsuperscript{5}, and Pa$\cdot$s\textsuperscript{5}, respectively).  Such a functional form, while not obvious, can be understood in the following way.  In the zero frequency limit, $ a = \eta_3^*( 0, 0, 0 ) $, and reflects the initial deviation from linearity in steady shear viscosity at small shear rates.  Values of $a < 0$ therefore indicate the onset of shear thinning in steady shear. The imaginary part of $\eta^*_3(\omega_1,\omega_2,\omega_3)$ has linear dependence on frequency because it is the lowest order odd function in frequency possessing Hermitian symmetry.  The coefficient $ b $ is associated with the elastic response measured in a parallel superposition experiment, again at small strain rate and strain amplitudes.  The coefficients $ c $ and $ d $ are associated with the only quadratic terms formed from the set of frequencies that respect permutation symmetry.

\begin{figure}[t]
    \centering
    \includegraphics[scale=0.6]{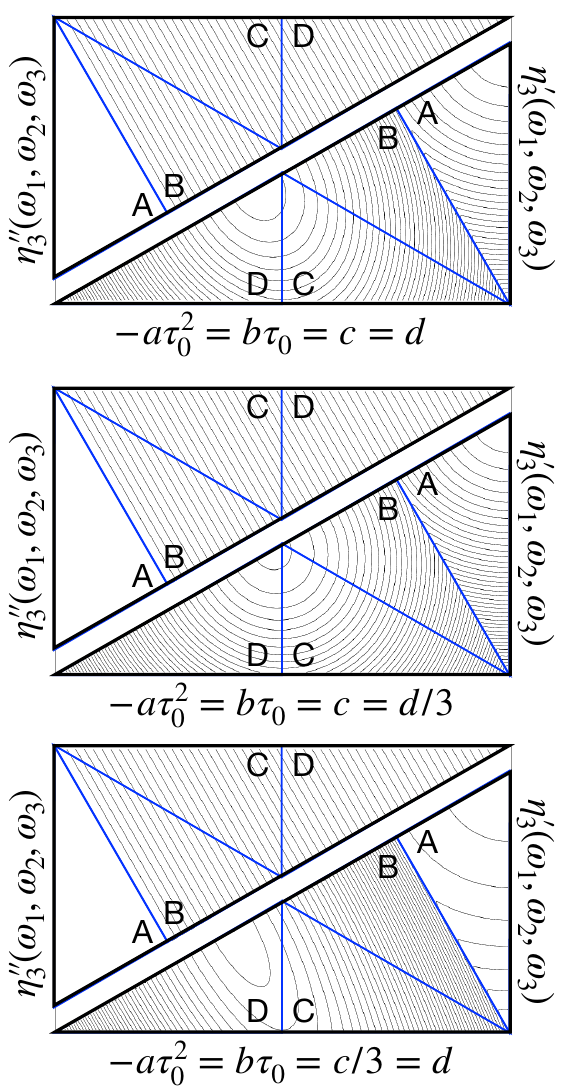}
    \caption{The third order complex viscosity in a viscoelastic fluid with a compact spectrum of relaxation times is quadratic as indicated in equation \ref{eq:lowfreq}.  Contours of this quadratic function are plotted in the MAPS triangles A, B, C, and D at $ \mathrm{De}_1 = 0.1 $ for different ratios of the polynomial coefficients. For compactness the imaginary component is shown inverted and rotated by $180^{\circ}$. The four triangles A, B,C, and D are each labelled.}
    \label{fig:low_freq}
\end{figure}

The low frequency limit in a viscoelastic fluid with longest relaxation time, $ \tau_0 $, can be identified by the value of the dimensionless group:
\begin{equation}
\mathrm{De}_1 = \tau_0 \| \boldsymbol{\omega} \|_1, \label{eq:De}
\end{equation}
which is the Deborah number based on the one-norm of the arguments to the third order complex viscosity.  This dimensionless group represents the product of the characteristic relaxation time in the fluid, $ \tau_0 $, with the maximum rate of change of a time varying strain signal containing three tones at equal amplitude with zero phase lag.  When $ \mathrm{De}_1 \ll 1 $, the flow changes more slowly than all the microstructural relaxation processes in the material. Figure \ref{fig:low_freq} plots contour maps of the real part of the complex viscosity in the triangles A, B, C, and D for different values of $ c / d $.  The imaginary part of the complex viscosity is simply a linear function of the frequencies.  On a constant $L^1$-norm surface, this gives linear contours.  The real part is quadratic, which gives ellipsoidal contours on each distinct face of the $L^1$-norm surface.  When $ c = 2 d $, this quadratic part can be written in terms of just the frequency sums as well: $ d ( \sum_j \omega_j )^2 $.  In such a case, transformation of the low frequency third order complex viscosity back to the time domain would give a third order  relaxation modulus that depends only on one single historic time. Equivalently, when $ c \ne 2 d $, the single historic time approximation is broken on time scales for which $\omega^{-1} < (d/a)^{1/2}$, when the quadratic terms in equation \ref{eq:lowfreq} become significant.

This functional form for the complex viscosity at low frequency also suggests the following behavior for MAOS material functions, obtained by substituting equation \ref{eq:lowfreq} into equations \ref{eq:e1} through \ref{eq:v3} from section \ref{sec:maos}, found by substituting equation \ref{eq:lowfreq} into equation \ref{eq:maos_all}:
\begin{subequations}
\begin{align}
    [v_1](\omega)  = \frac{3\omega^2}{4} \left( a + ( 3 d - c ) \omega^2 \right), \\
    [v_3](\omega)  = \frac{\omega^2}{4} \left( a + 3 ( d + c ) \omega^2 \right), \\
    [e_1](\omega ) = -[e_3](\omega) = \frac{3 b \omega^4}{4},
\end{align}
\end{subequations}
Evidence of the $ O( \omega^2 ) $ scaling for the viscous functions and $ O( \omega^ 4 ) $ for the elastic functions has been produced already in MAOS experiments \cite{ewoldt-2014}.  Therefore, the coefficients $ a $ and $ b $ are directly measurable via MAOS.  In principle, $ c $ and $ d $ could be determined by fitting MAOS data at low frequency, or by measuring the quantities $[v_1](\omega) - 3[v_3](\omega) = -3 c \omega^4 $ and $ [v_1](\omega) + [v_3](\omega) = a \omega^2 + 3 d \omega^4$, but it is the present MAPS framework that puts these coefficients in their proper context.

\subsubsection{Algebraic Constitutive Model: Generalized Newtonian Fluid}

Having developed the Volterra representation for MAPS strain-controlled rheology, we are now in a position to apply this formulation to viscoelastic constitutive models. While the representation is applicable to the simple shear response of models regardless of complexity, we can demonstrate the elegance of the representation by considering the simple case of a generalized Newtonian fluid, $ \sigma = \eta(\dot{\gamma}^2)\dot{\gamma} $ \cite{larson-1999}.
This is perhaps the simplest phenomenological model with a shear stress that depends nonlinearly on the deformation rate.  The shear stress can be expanded about small shear rates as:
\begin{equation}
    \sigma(t) = \eta(0) \dot{\gamma}( t ) + \eta^\prime( 0 ) \dot{\gamma}(t)^3  + O(\dot{\gamma}^5).
\end{equation}
Here, $\eta'(0)$ refers to the zero-shear limit of the first derivative of the shear rate-dependant viscosity function: $\eta'(0) = d\eta(\dot{\gamma}^2)/d(\dot{\gamma}^2)\rvert_{\dot{\gamma}^2 = 0}$. To third order, the Fourier transform of the stress is then:
\begin{equation}
    \hat{\sigma}(\omega) = \eta(0) \hat{\dot{\gamma}}( \omega ) + \frac{\eta^\prime(0)}{4\pi^2} \hat{\dot{\gamma}}( \omega )*\left(\hat{\dot{\gamma}}( \omega )*\hat{\dot{\gamma}}( \omega )\right),
\end{equation}
where $ \hat f( \omega ) * \hat g( \omega ) $ indicates a convolution.  The convolution formula:
\begin{align}
    &\hat{\dot{\gamma}}(\omega)*\left(\hat{\dot{\gamma}}(\omega)*\hat{\dot{\gamma}}(\omega)\right)
    \label{eq:conv_2} \\
    &= \iiint_{-\infty}^{\infty}\delta(\omega - \sum_j \omega_j)\hat{\dot{\gamma}}(\omega_1)\hat{\dot{\gamma}}(\omega_2)\hat{\dot{\gamma}}(\omega_3)\, d\omega_1 d\omega_2 d\omega_3. \nonumber
\end{align}
can be used to rewrite the shear stress in the form of a Volterra kernel expansion.  By comparison of the Fourier transformation of the shear stress with equation \ref{eq:strain_controlled3}, the first and third order complex viscosities for a generalized Newtonian fluid can be identified directly: $\eta^*_1(\omega) = \eta(0) $, and $ \eta^*_3(\omega_1,\omega_2,\omega_3) = \eta^\prime(0) $.

The simplicity of the expressions for $\eta^*_1( \omega )$ and $\eta^*_3( \omega_1, \omega_2, \omega_3 )$ for this canonical inelastic model indicate that the formulation we have developed is a natural representation for the weakly nonlinear response to simple shearing deformation. In fact, because the rheological response of this simple material is instantaneous in time (i.e. there is no memory of the deformation history), the $n$th order complex viscosities are exactly the same as the Taylor coefficients for the expansion of the viscosity function, $ \eta( \dot \gamma^2 ) $ with respect to $ \dot \gamma^2 $. Because these coefficients are real-valued, the first and third-order complex viscosities have only real components, as expected for an inelastic fluid.  Consider for example, the Carreau model, for which
$ \eta(\dot{\gamma}^2) = \eta_{\infty} + (\eta_0 - \eta_{\infty})(1 + (\lambda\dot{\gamma})^2)^\frac{n-1}{2} $.  Then, the first and third order complex viscosities are simply $\eta^*_1(\omega) = \eta_0 $ and $ \eta^*_3(\omega_1,\omega_2,\omega_3) = ( n - 1 ) (\eta_0 - \eta_{\infty})\lambda^2 / 2 $.  In the Carreau model, when $ n < 1 $, and $ \eta_0 > \eta_\infty $ the fluid is shear thinning.  We see here that the same conditions result in $ \eta^*_3( \omega_1, \omega_2, \omega_3 ) < 0 $.  Thus, a negative real part of the third order complex viscosity is associated with a shear thinning response.

\subsubsection{Differential Constitutive Model: Simple Fluidity}

To capture time-dependent nonlinear viscoelastic behavior, we examine a simple fluidity model \cite{moorcroft-2011}:
\begin{equation}
    \frac{d\sigma}{dt} = G\dot{\gamma}(t) - \frac{1}{\tau_0}f(\dot{\gamma}(t)^2)\sigma(t), \label{eq:fluidity}
\end{equation}
in which changes in shear stress are driven both elastically: i.e. by $ G \dot \gamma( t ) $, and by relaxation: $ - \tau_0^{-1} f(\dot{\gamma}(t)^2)\sigma(t) $, with a relaxation rate that depends nonlinearly on the current shear rate.  In this simple fluidity model, the leading order nonlinearities are prescribed by the functional form of $f(\dot{\gamma}(t)^2)$:
$f(\dot{\gamma}^2) = 1 + f^\prime( 0 ) \dot{\gamma}^2 + O(\dot{\gamma}^4)$,
for which we take $ f( 0 ) = 1 $ without loss of generality.
To third order, the Fourier transformation of equation \ref{eq:fluidity} then gives:
\begin{align}
    & i\omega\hat{\sigma}(\omega) \\
    &\quad = G\hat{\dot \gamma}(\omega) - \frac{1}{\tau_0}\left(\hat{\sigma}(\omega) + \frac{f^\prime( 0 )}{4\pi^2} \hat{\dot \gamma}(\omega) * \hat{\dot \gamma}(\omega)* \hat{\sigma}(\omega) \right). \nonumber
\end{align}
The contributions to the $ n $th order complex viscosities can be identified by substituting equation \ref{eq:strain_controlled3} for the third order Volterra expansion of $\hat{\sigma}(\omega)$ into this expression and then grouping terms of the same order of magnitude in $ \hat{ \dot \gamma}( \omega ) $.  At first order, one finds that:
\begin{equation}
    i\omega\eta^*_1(\omega)\hat{\dot \gamma}(\omega) = G\hat{\dot \gamma}(\omega) - \frac{1}{\tau_0}\eta^*_1(\omega)\hat{\dot \gamma}(\omega),
\end{equation}
which can be solved for the first order complex viscosity,
\begin{equation}
    \eta^*_1(\omega) = \frac{G\tau_0}{1 + i\tau_0\omega}.
\end{equation}
Written in terms of its real and imaginary parts, $\eta^*_1(\omega) = \eta'_1(\omega) - i\eta''_1(\omega)$, this gives:
\begin{equation}
    \eta'_1(\omega) = \frac{G\tau_0}{1 + \tau_0^2\omega^2}, \quad \eta''_1(\omega) = \frac{G\tau_0^2\omega}{1 + \tau_0^2\omega^2}.
\end{equation}
The first order complex viscosity thus takes the familiar form of a single Maxwell mode, consistent with the Maxwellian character of the generalized fluidity model.  At third order, we find
\begin{align}
    & \iiint_{-\infty}^{\infty}\eta^*_3(\omega_1,\omega_2,\omega_3) \delta(\omega-\sum_j \omega_j) \\ 
    & \quad\quad\quad\quad\quad \times \hat{\dot \gamma}(\omega_1)\hat{\dot \gamma}(\omega_2)\hat{\dot \gamma}(\omega_3) \, d\omega_1d\omega_2d\omega_3 \nonumber \\ & \quad= -f^\prime(0) \left(\frac{1}{1 + i\tau_0\omega}\right)\left\{\hat{\dot \gamma}(\omega)*\hat{\dot \gamma}(\omega)*\left[\eta^*_1(\omega)\hat{\dot \gamma}(\omega)\right]\right\}. \nonumber
\end{align}
By analogy to Equation \ref{eq:conv_2}, we can write the triple convolution as:
\begin{align}
    & \hat{\dot \gamma}(\omega)*\hat{\dot \gamma}(\omega)*\left[\eta^*_1(\omega)\hat{\dot \gamma}(\omega)\right] \\
    & \quad = \iiint_{-\infty}^{\infty} \eta^*_1(\omega_1) \delta(\omega - \sum_j \omega_j)  \nonumber \\ 
    & \quad \quad \quad \quad \quad \quad \times \hat{\dot \gamma}(\omega_1) \hat{\dot \gamma}(\omega_2)\hat{\dot \gamma}(\omega_3)\,d\omega_1d\omega_2d\omega_3. \nonumber
\end{align}
Making use of the sifting property of the delta function allows one to move the factor: $1/(1 + i\tau_0\omega)$, inside the convolution integrals with the substitution: $\omega \rightarrow \sum_j \omega_j $. Thus, we find that the third order complex viscosity for the generalized fluidity model is:
\begin{equation}
    \eta^*_3(\omega_1,\omega_2,\omega_3) = -f^\prime(0) \left[\frac{\eta^*_1(\omega_1)}{1 + i\tau_0 \sum_j \omega_j }\right].
\end{equation}
However, one should immediately notice that this expression does not respect permutation symmetry. This is because of the arbitrary choice to associate the term: $\eta^*_1(\omega)$, in the convolutions with frequency $\omega_1$. Such lapses of symmetry are easily repaired.  Here, we simply average over the three possible choices in rewriting the convolutions, giving a permutation-symmetric expression for the third order complex viscosity,
\begin{align}
    & \eta^*_3(\omega_1,\omega_2,\omega_3) \\
    &= -\frac{G\tau_0}{3} f^\prime(0) \left(\frac{1}{1 + i\tau_0 \sum_j \omega_j }\right)\sum_j\frac{1}{1 + i\tau_0\omega_j} \nonumber\\
    & = -\frac{1}{3} \frac{f^\prime( 0 )}{G \tau_0} \eta^*_1( \omega_1 +\omega_2 + \omega_3 ) \sum_j \eta^*_1(\omega_j), \nonumber
\end{align}
that can be written in terms of the first order complex viscosity and the unknown material property $f'(0)$ alone.

Finally, we note that the simple fluidity model possesses only a single relaxation time.  Therefore, in the low-frequency limit, its third order complex viscosity can be written as a quadratic function in frequency, in the form of equation \ref{eq:lowfreq}.  The set of four scalar coefficients, defined in Section \ref{sec:lowfreq}, representing modes of this polynomial that are consistent with the symmetries of the response function are:
\begin{equation}
    \left( \begin{array}{c}a\\b\\c\\d\end{array}\right) = \frac{G \tau_0 f^\prime( 0 )}{ 3} \left( \begin{array}{c}-1 \\ 4 \tau_0 \\ 8 \tau_0^2\\ 5 \tau_0^2 \end{array} \right). \label{eq:lowfreqfluidity}
\end{equation}

\begin{figure*}[t]
    \centering
    \includegraphics[width=1.6\columnwidth]{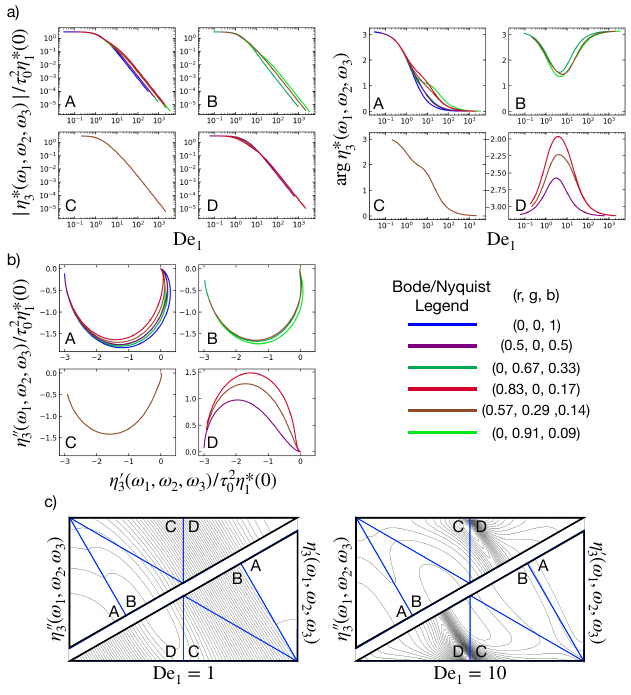}
    \caption{MAPS rheology for the fluidity model with $f'(0) = 3 \tau_0^2 $. a) Bode plots of the magnitude and phase as a function of the $L^1$-norm frequency in different MAPS triangles. b) Nyquist plots in different MAPS triangles. c) Contour plots at different $ \mathrm{De}_1 $ on the constant $L^1$-norm surface.}
    \label{fig:fluidity}
\end{figure*}

Figure \ref{fig:fluidity}a) and \ref{fig:fluidity}b) depict Nyquist and Bode plots of the third order complex viscosity for the simple fluidity model at some select barycentric coordinates in the A, B, C, D subspaces. In the Bode plot, the magnitude of the complex viscosity and the phase angle are plotted against $ \mathrm{De}_1 $ (defined in equation \ref{eq:De}), using the relaxation time of the fluidity model, $ \tau_0 $ to make the $L^1$ norm of the frequency dimensionless.  Finally, in Figure \ref{fig:fluidity}d) we plot a contour map of the real and imaginary parts of the complex viscosity across the connected set of triangular subspaces at two different values of $ \mathrm{De}_1 $.  At small values of $ \mathrm{De}_1 \ll 1 $, the complex viscosity is consistent with the quadratic low-frequency expansion in equation \ref{eq:lowfreq}; thus its imaginary and real components have linear and elliptical contours, respectively, when projected onto the triangles as shown in Figure \ref{fig:low_freq}.  At high frequency, the magnitude of the third order complex viscosity decays as $ \mathrm{De}_1^{-2} $, which, as expected from the analysis of MAOS at high frequencies, is faster than the $ \mathrm{De}_1^{-1} $ decay in the linear response \cite{swan-2014,swan-2016}.


\subsubsection{Integral Constitutive Model: Lodge-like Fluid}

In the generalized Newtonian and simple fluidity models, we considered a simple algebraic model and a simple differential model, respectively. Many viscoelastic constitutive models, however, are expressed in terms of as integrals. One simple example of a generalized integral model is a Lodge-like model, which when written in scalar form for the shear stress reads \cite{vermant-1998}:
\begin{align}
    \sigma(t) &= \int_{-\infty}^{t} \int_{-\infty}^{\infty}\frac{1}{\tau} H(\tau, 2 \dot{ \gamma }( t^\prime ) ^ 2) e^{-(t-t')/\tau} d\ln\tau \nonumber \\ 
    & \quad\quad\quad\quad\quad\quad\times (\gamma( t ) - \gamma( t^\prime ) ) \, dt',
\end{align}
where $ H( \tau, 2 \dot{ \gamma }( t^\prime ) ^ 2 ) $ is a relaxation time distribution that depends parametrically on the second invariant of the deformation rate tensor: $ 2 \dot{ \gamma }( t^\prime ) ^ 2 $, in simple shear flow. To describe the weakly nonlinear response of this model, the relaxation spectrum can be expanded at small strain,
\begin{align}
    H(\tau, 2 \dot{ \gamma }^2 ) = H_0(\tau)\left(1 + \dot{ \gamma }^2 H_1(\tau) \right) + O(\dot{ \gamma }^4) \nonumber
\end{align}
with $ H_0(\tau) = H(\tau,0) $,  $ H_1(\tau) = \left. \frac{d \ln H}{d (\dot{ \gamma }^2)}\right|_{\dot{ \gamma } = 0} $. Defining the expansion in this way is convenient because it allows for a factor of $ H_0(\tau) $, the equilibrium relaxation spectrum, to be separated from the strain rate-dependent quantity $( 1 + \dot{\gamma}^2 H_1(\tau) )$. As we demonstrate below, this simplifies the notation by permitting a consistent definition of an averaged quantity.

To leading order in the shear strain or strain rate, the shear stress is:
\begin{align}
    \sigma(t) &= \int_{-\infty}^{t} \int_{-\infty}^{\infty}\frac{1}{\tau} H_0(\tau) e^{-(t-t')/\tau}d\ln\tau (\gamma(t) - \gamma(t'))dt'  \nonumber \\
    &=  \int_{-\infty}^{t} \left< \frac{1}{\tau}e^{-(t-t')/\tau}\right>(\gamma(t) - \gamma(t')) \, dt' ,
\end{align}
where the angle brackets represent an average of the argument over the equilibrium relaxation time spectrum, i.e.: 
\begin{equation}
    \langle x(\tau) \rangle = \int_{-\infty}^{\infty}x(\tau)H_0(\tau) \, d\ln\tau.
    \label{eq:lodge_average}
\end{equation}
Making the change of variables: $ u = t - t'$, and representing the strain in terms of its Fourier transformation gives:
\begin{align}
    \sigma(t) &=
    \frac{1}{2\pi}\int_{-\infty}^{\infty} e^{i\omega t} \hat \gamma( \omega ) \\
    & \quad\quad\quad \times \left< \frac{1}{\tau} \left(\int_{0}^{\infty}e^{-u/\tau}(1-e^{-i\omega u}) du \right) \right> \,  d\omega. \nonumber
\end{align}
Integration over the dummy variable $ u $, allows for direct identification of the linear response:
\begin{equation}
    G^*_1(\omega) = \left\langle \frac{i\omega\tau}{1+i\omega\tau} \right\rangle.
    \label{eq:lodge_G1}
\end{equation}
As expected of a Lodge-like model, the first order complex modulus is an average of Maxwell modes taken over the equilibrium relaxation time distribution.

To $O(\gamma(t)^3)$, the shear stress is:
\begin{align}
    \sigma(t) &= \frac{1}{2\pi}\int_{-\infty}^\infty e^{i \omega t} G_1^*( \omega ) \hat \gamma( \omega ) \, d \omega  \\ &+  \left< \frac{H_1(\tau)}{\tau}  \int_{0}^{\infty}  e^{-u/\tau} \dot{ \gamma }(t-u)^2 (\gamma(t) - \gamma(t-u)) \, d u \right>. \nonumber
\end{align}
Rewriting the stress, strain and strain rate in terms of their Fourier transformations and integrating over the dummy variable $ u $ allows for identification of the third order complex modulus by comparison with equation \ref{eq:strain_controlled3}:
\begin{align}
    \hat \sigma(\omega) &= G_1^*( \omega ) \hat \gamma( \omega ) \\
    & + 
    \frac{1}{4\pi^2}\iiint_{-\infty}^\infty \delta( \omega - \sum_j \omega_j ) \, \omega_2 \omega_3 \nonumber\\ & \times \left< H_1( \tau ) \left( \frac{1}{1+i(\sum_j \omega_j)\tau} -  \frac{1}{1+i(\omega_2+\omega_3)\tau}\right)
    \right> \nonumber \\
    & \times\hat{\gamma}(\omega_1)\hat{\gamma}(\omega_2)\hat{\gamma}(\omega_3) \, d\omega_1d\omega_2d\omega_3, \nonumber
\end{align}
which upon permutation symmetrization can be represented as:
\begin{align}
     &G^*_3(\omega_1,\omega_2,\omega_3) = \label{eq:lodgemodulus} \\ &\frac{1}{3} \sum_{k\neq j} \left[\left\langle\frac{H_1(\tau)\omega_j\omega_k}{1+i\tau\sum_{l}\omega_l}\right\rangle - \left\langle\frac{H_1(\tau)\omega_j\omega_k}{1+i\tau(\omega_j+\omega_k)}\right\rangle\right]. \nonumber
\end{align}
Thus, the third order complex viscosity in the Lodge-like model is:
\begin{align}
    &\eta^*_3(\omega_1,\omega_2,\omega_3) = \label{eq:lodgevisc}\\ &\frac{i}{3 \omega_1\omega_2\omega_3} \sum_{k\neq j} \left[\left\langle\frac{H_1(\tau)\omega_j\omega_k}{1+i\tau\sum_{l}\omega_l}\right\rangle - \left\langle\frac{H_1(\tau)\omega_j\omega_k}{1+i\tau(\omega_j+\omega_k)}\right\rangle\right]. \nonumber
\end{align}
Notably, $\eta^*_3( \omega_1, \omega_2, \omega_3 ) $ for the general Lodge-like model cannot be expressed in terms of $\eta^*_1( \omega ) $ (or $G^*_1(\omega)$), unlike for the fluidity model.  In the low frequency limit, the third order complex viscosity for a Lodge-like fluid with a compact relaxation time spectrum, $\log H(\tau, 2\dot{\gamma}(t')^2) \sim -\tau^p$ with $p > 0$, can again be written as a quadratic function of the frequency.  The four scalar coefficients defined in Section \ref{sec:lowfreq} for the Lodge-like model are:
\begin{equation}
    \left( \begin{array}{c} a \\ b \\ c\\ d \end{array} \right) = -\left< \frac{ H_1( \tau ) \tau }{3} \left( \begin{array}{c} -3 \\ 5 \tau \\ 12 \tau^2 \\ 7 \tau^2 \end{array} \right) \right>. \label{eq:lowfreqlodge}
\end{equation}
Although the functional form of the low frequency modulus is similar to equation \ref{eq:lowfreqfluidity}, ratios of the scalar coefficients depend here on moments of the perturbed relaxation time distribution.  

When the Lodge-like fluid possesses only a single relaxation time, $ \tau_0 $, the relaxation time distribution can be written as $H(\tau) = \tau_0 \delta(\tau - \tau_0)$ and the definition of an average over $H(\tau)$ takes on a simple form:
\begin{equation}
    \langle x(\tau) \rangle = x(\tau_0).
\end{equation}
The low frequency expansion coefficients can thus be determined by dropping the angle brackets in equation \ref{eq:lowfreqlodge} and replacing $ \tau $ with $ \tau_0 $.  In this form it is clear that the low frequency moduli of the weakly nonlinear Lodge-like constitutive model are quite distinct from that of the fluidity model.  We expect this to hold in general across the wide spectrum of models for non-Newtonian fluids.  Identification of the scalar coefficients $ a, b, c, d $ should thus be a principal objective when characterizing nonlinearities in complex fluids.  We also note that in the absense of a compact relaxation time distribution, as with gels possessing power-law relaxation spectra, equation \ref{eq:lowfreq} will not hold in general and anomalous power law dependence of third order moduli on frequency are anticipated at low frequency.  Consideration of such materials and their nonlinear response is left to future work.

\begin{figure*}[t]
    \centering
    \includegraphics[width=1.6\columnwidth]{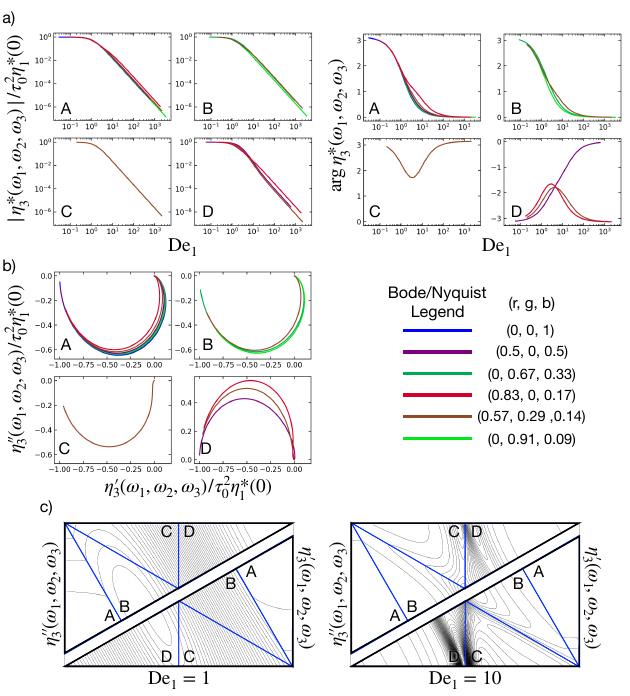}
    \caption{MAPS rheology for a single mode Lodge-like model with $ H_1( \tau_0 ) = \tau_0^2 $. a) Bode plots of the magnitude and phase as a function of the $L^1$-norm frequency in different MAPS triangles.  b) Nyquist plots in different MAPS triangles.  c) Contour plots at different $ \mathrm{De}_1 $ on the constant $L^1$-norm surface.}
    \label{fig:lodge}
\end{figure*}

Figure \ref{fig:lodge}a) depicts a Bode plot of the third order complex viscosity for a single mode Lodge-like model  at some select barycentric coordinates in the four triangular MAPS regions: A, B, C, and D.  As with the fluidity model, the magnitude of the complex viscosity decays as  $ \| \boldsymbol{\omega} \|_1^{-2} $ at high frequency.  The phase behavior of the Lodge model, on the other hand, is quite distinct from that of the fluidity model. Figure \ref{fig:lodge}b) is a Nyquist plot of the same data.  Figure \ref{fig:lodge}c) plots the contours of the third order complex viscosity across the connected triangular subspaces at two different values of $ \mathrm{De}_1 $.  The pattern of contours is also quite clearly distinct from that of the simple fluidity model shown in Figure \ref{fig:fluidity}c).

One intriguing application of a MAPS data set is in the reconstruction of $ H_1( \tau ) $ from the measured response.  Just as the linear response given by equation \ref{eq:lodge_G1} can be inverted via equation \ref{eq:lodge_average} to infer the equilibrium relaxation time distribution, $ H_0( \tau ) $, equations \ref{eq:lodgemodulus} or \ref{eq:lodgevisc} could be inverted to approximate the dependence of the relaxation time distribution on shear rate, $ H_1( \tau ) $.  Because the MAPS data is drawn from a high dimensional manifold, any estimate of $ H_1( \tau ) $ that fits the measured data well and uniformly should be considered strong evidence that the Lodge-like model provides a sound description of the measured nonlinear viscoelasticity.  Identification of this model using data from a lower dimensional manifold such as MAOS will provide much weaker evidence because the underlying third order moduli change in distinct ways away from the vertices in each of the four unique MAPS subspaces.  Even though higher order dependencies of the relaxation time distribution on shear rate are not measured via the weakly nonlinear MAPS rheology, the leading order nonlinearity can be used to create an approximation for the full distribution.  Such schemes have proven useful for describing the nonlinear rheology of polymeric materials \cite{carreau-1968}.  Two such approximations which give the correct MAPS response are:
\begin{align}
    H( \tau, \dot \gamma^2 ) &\approx \frac{C_1( \dot \gamma ) H_0( \tau )}{ 1 - H_1( \tau ) \dot \gamma^2} \\ &\approx C_2( \dot \gamma ) H_0( \tau ) \exp \left( H_1( \tau ) \dot \gamma^2 \right), \nonumber
\end{align}
where the normalizing functions $ C_1( \dot \gamma ) $ and $ C_2( \dot \gamma ) $ vary as $ 1 + O( \dot \gamma^4 ) $, thus are constant at leading order in $ \dot \gamma $ and serve to properly normalize the relaxation time distribution.

\subsubsection{Time-Strain Separable Constitutive Model}
Constitutive models of nonlinear viscoelasticity based on the principle of time-strain separability can be written in integral form as:
\begin{equation}
    \sigma( t ) = \int_{-\infty}^t m( t - t^\prime ) \gamma( t, t^\prime ) h( \gamma( t, t^\prime )^2 ) d t^\prime,
\end{equation}
where $ m( t - t^\prime ) $ is the linear response memory kernel and given by the time derivative of the relaxation modulus.  The damping function $ h( \gamma( t, t^\prime ) ^ 2 ) $ is an even function of the accumulated strain between times $ t $ and $ t^\prime $: ${ \gamma( t, t^\prime ) = \gamma( t ) - \gamma( t^\prime )} $.

The most notable example of this sort of approximation is the factorized K-BKZ constitutive model \cite{bernstein-1963}.  For these sorts of models, the $ n $th order complex moduli can be written strictly in terms of the linear response.  The process proceeds much as for the generalized Newtonian constitutive model.  The damping function is expanded as a power series in the square of the accumulated strain, and the Fourier transformation of the stress and stress is substituted.  For the third order complex modulus of a strain separable constitutive model, one finds that:
\begin{align}
& G_3^*( \omega_1, \omega_2, \omega_3 ) = \left. \frac{\partial h( \gamma ^ 2 ) }{\partial ( \gamma^2 )} \right|_{\gamma = 0} \left[ G_1^*\left( \sum_{j=1}^3 \omega_j \right) \right. \label{eq:tss} \\
& \left. \quad  - \sum_{j=1}^3 G_1^*\left( \sum_{\substack{k=1 \\ k\ne j}}^3 \omega_k \right) + \sum_{j=1}^3  G_1^*( \omega_j ) - G_1^*( 0 ) \right]. \nonumber
\end{align}
This is a necessary but not sufficient condition to identify strain separability and is a generalization of expressions derived by Martinetti and Ewoldt for MAOS of strain separable models \cite{ewoldt-2019}. It is worth noting that the MAPS response of a time-strain separable model does not depend on the details of the damping function except for the asymptotic value of its first derivative with respect to $\gamma^2$.

Analogous expressions for the higher order complex moduli written in terms of the linear response can be easily derived.  The $n$th order modulus, $ G_n^*( \omega_1, \ldots \omega_n ) $ is a linear combination of $ G_1^*( \omega ) $ evaluated at all possible frequencies, $ \omega $, given by a sum over subsets of $ \{ \omega_1, \ldots, \omega_n \} $ including the null set.  The amplitudes assigned to the terms in this linear combination are:
\begin{equation}
    (-1)^{m+1}  \left. \frac{\partial^{(n-1)/2} h( \gamma^2 ) }{ \partial ( \gamma^2 )^{(n-1)/2}} \right|_{\dot \gamma = 0},
\end{equation} 
where $ m $ is the number of terms in each subset of frequencies.

For a time-strain separable fluid, equation \ref{eq:tss} dictates that the functional forms of the third order complex modulus $G^*_3( \omega_1, \omega_2, \omega_3 )$ and viscosity $\eta^*_3( \omega_1, \omega_2, \omega_3 ) $ are determined entirely by the linear response function $G^*_1( \omega )$. For a family of constitutive models with a common linear response function, then, there is only a single form of any third order MAPS response function that obeys time-strain separability. This principle can be used to easily determine whether any specific constitutive model is time-strain separable.

\begin{figure*}[t]
    \centering
    \includegraphics[width=1.6\columnwidth]{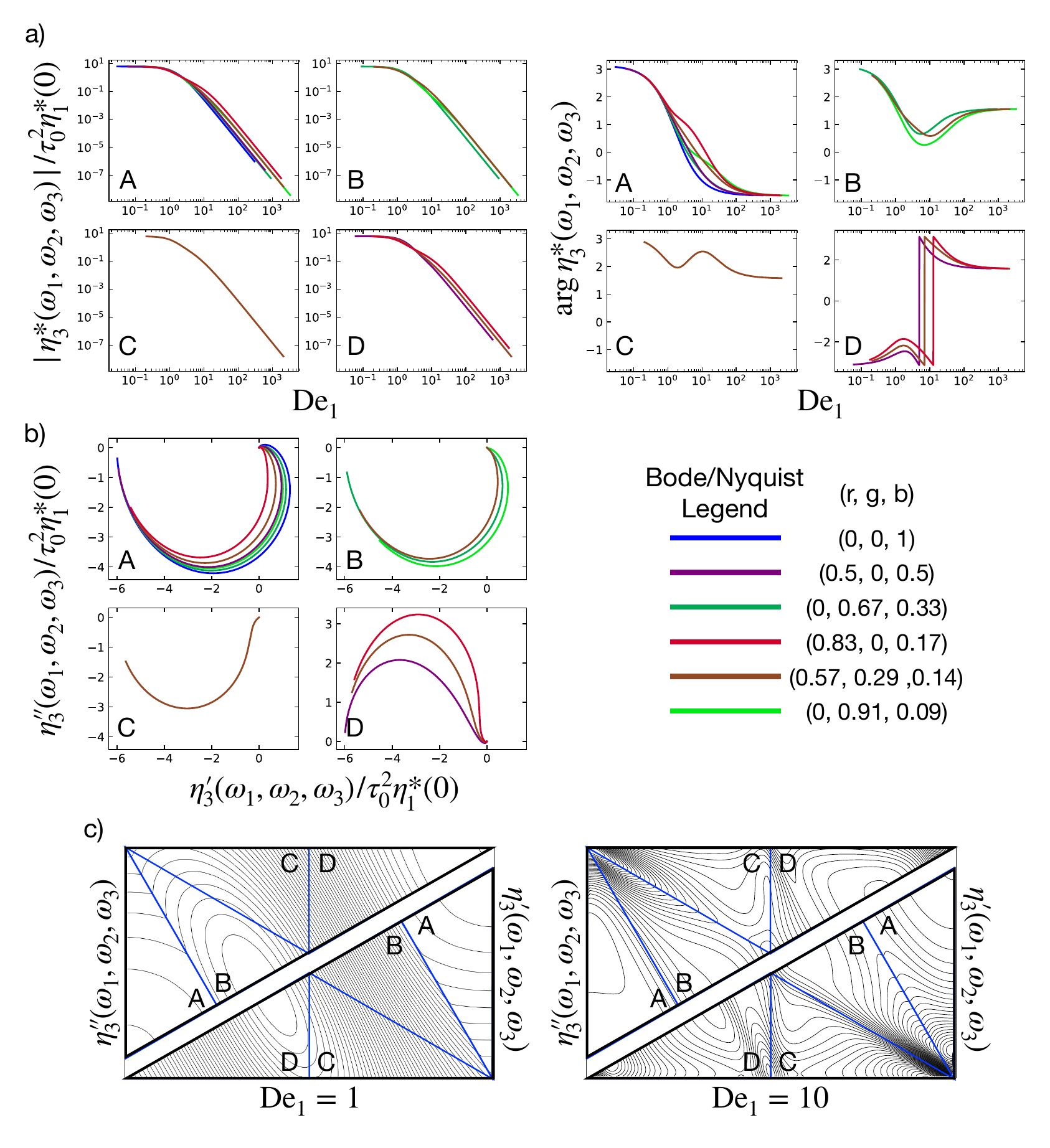}
    \caption{MAPS rheology for the time-strain separable Maxwell model with $ \frac{\partial h(\gamma^2)}{\partial(\gamma^2)}\rvert_{\gamma=0} = -1 $. a) Bode plots of the magnitude and phase as a function of the $L^1$-norm frequency in different MAPS triangles.  b) Nyquist plots in different MAPS triangles.  c) Contour plots at different $ \mathrm{De}_1 $ on the constant $L^1$-norm surface.}
    \label{fig:tss}
\end{figure*}

Both the generalized fluidity and single relaxation time Lodge-like models presented above share a common linear response function: that of a single Maxwell mode. It is clear from detailed comparison of Figures \ref{fig:fluidity} and \ref{fig:lodge}, however, that their third order complex viscosities are quite different. Thus both models cannot be time-strain separable. To check whether either model is time-strain separable, we can construct the solution for $\eta^*_3(\omega_1,\omega_2,\omega_3)$ using equation \ref{eq:tss} in conjunction with equation \ref{eq:G_to_eta} for the Maxwell fluid: $ G_1^*(\omega) = G i \omega \tau_0 / ( 1 + i \omega \tau_0 ) $. Bode and Nyquist diagrams of this solution (consistent with time-strain separability), as well as contour plots on two constant $L^1$-norm surfaces, are presented in Figure \ref{fig:tss}.

Martinetti and Ewoldt have shown that multiple models belong to the class of time-strain separable single-mode Maxwell models, including the corotational Maxwell model, and both the linear and quadratic molecular stress function models \cite{ewoldt-2019}. In Appendix \ref{app:crm}, we prove this for the corotational Maxwell model by solving for $\eta^*_3(\omega_1,\omega_2,\omega_3)$ directly from the model and by using equation \ref{eq:tss}. Though there are numerous examples of time-strain separable single-mode Maxwell models, it is clear that the curves and contour plots in Figure \ref{fig:tss} are distinct from those in Figures \ref{fig:fluidity} and \ref{fig:lodge} for the generalized fluidity and Lodge models, respectively. Therefore, \textit{neither} of these two classes of models obeys time-strain separability.

The assumption of time-strain separability is convenient for modeling weakly nonlinear rheology because it requires the measurement of only a single nonlinear parameter, the first derivative of the damping function $ h^\prime(0) $, in addition to the linear response function $G^*_1(\omega)$. In a sense, it is the simplest nonlinear constitutive relation possible, in terms of information content about a material's nonlinear response. A simple first test to perform on MAPS data whose underlying structure is unknown is to compare it to the time-strain separable prediction given the material's linear response, which will indicate whether the time-strain separable assumption is valid.  A comparison of Figures \ref{fig:fluidity} and \ref{fig:lodge} with Figure \ref{fig:tss}, for example, reveals that such an assumption can be both quantitatively and qualitatively inaccurate for certain classes of fluids or constitutive models. The time-strain separable single-mode Maxwell solution predicts that $|\eta^*_3( \omega_1, \omega_2, \omega_3 )|$ should scale with $\text{De}_1^{-3}$ at high frequencies, while both the fluidity and Lodge models predict a high frequency scaling of $\text{De}_1^{-2}$. The time-strain separable solution also predicts that $\text{arg} \left(\eta^*_3( \omega_1, \omega_2, \omega_3 ) \right)$ should approach asymptotic limits of integer multiples of $\pi$ at low frequency, and half-integer multiples of $\pi$ at high frequency. In other words, it predicts that $\eta^*_3( \omega_1, \omega_2, \omega_3 ) $ should be dominated by $\eta'_3( \omega_1, \omega_2, \omega_3 ) $, or viscous nonlinearities, at low frequencies, but by $\eta''_3( \omega_1, \omega_2, \omega_3 )$, or elastic nonlinearities, at high frequencies. While both the fluidity and Lodge models have the same character at low frequencies, they predict dominance by $\eta'_3(\omega_1,\omega_2,\omega_3)$, or viscous nonlinearities, at high frequencies instead. Thus the time-strain separable solution fails to capture features of the fluidity and Lodge models that substantially impact their physical interpretation.

In the low frequency limit, the third order complex viscosity of a fluid that demonstrates time-strain separability and for which the linear response is given by a single Maxwell mode has the expected low frequency expansion with coefficients:
\begin{equation}
    \left( \begin{array}{c} a \\ b \\ c\\ d \end{array} \right) = -2 G \tau_0^3 h^\prime( 0 )  \left( \begin{array}{c} -3 \\ 6 \tau_0 \\ 15 \tau_0^2  \\ 10 \tau_0^2 \end{array} \right). \label{eq:lowfreqtss}
\end{equation}
The ratios $b/a$, $c/a$, and $d/a$ obviously differ from any of the other models discussed.

\subsubsection{Tensorial Constitutive Model: Giesekus Fluid}
\label{sec:giesekus}

In general, stresses in flows of complex fluids are tensorial in nature, and are represented by the tensor $\boldsymbol{\sigma}$. The off-diagonal elements of $\boldsymbol{\sigma}$ represent shear stresses, and the diagonal elements represent normal stresses. To account for the tensorial nature of stresses in viscoelasticity, many constitutive models are written in a tensorial form. One such constitutive model that has found widespread utility is the Giesekus model \cite{giesekus-1982}:
\begin{equation}
    \boldsymbol{\sigma} + \tau_0 \boldsymbol{\sigma}_{(1)} + \frac{\alpha \tau_0}{\eta_0}\boldsymbol{\sigma}\cdot\boldsymbol{\sigma} = \eta_0\boldsymbol{\dot{\gamma}}, \label{eq:giesekus}
\end{equation}
where $\boldsymbol{\sigma}_{(1)}$ represents the upper convected derivative of the stress tensor, and $\boldsymbol{\dot{\gamma}}$ is the rate-of-strain tensor. The Giesekus model can be derived from a network theory of polymer melts or for a dilute suspension of dumbbells with anisotropic drag \cite{bird-1987-2}. The stress tensor $\boldsymbol{\sigma}$ in equation \ref{eq:giesekus} represents the polymer contribution to the stress tensor, which may be superimposed on a Newtonian solvent contribution. The model is parameterized by a zero-shear viscosity $\eta_0$, a characteristic relaxation time $\tau_0$, and a mobility parameter $\alpha$, which originates from hydrodynamic interactions between polymer chains \cite{bird-1987}.

The linear response of a Giesekus fluid is identical to that of the previously considered generalized fluidity and Lodge-like models - a single-mode Maxwell response $\eta^*_1(\omega) = \eta_0/(1 + i\omega\tau_0)$. Due to the tensorial nature of the model, the derivation of the third order complex viscosity is more involved than for previously considered models, thus we omit the mathematical details here for brevity, but provide a derivation in Appendix \ref{app:giesekus}. The solution involves expanding $\boldsymbol{\sigma}$ as a power series in the characteristic amplitude of the MAPS flow, $\gamma_0$, then performing asymptotic matching and Fourier transformations to sequentially solve for the components of $\boldsymbol{\sigma}$ at increasing order. After simplification, the solution for the third order complex viscosity is:
\begin{align}
    & \frac{ \eta^*_3( \omega_1, \omega_2, \omega_3 ) }{ \eta_0 \tau_0^2 } = \frac{ \alpha \left( (3 - 2\alpha) + i\tau_0 \sum_j\omega_j \right) }{ 3 \left(\prod_j(1 + i\tau_0\omega_j)\right) } \times \label{eq:giesekus_eta3} \\
    & \frac{ \left(-3 - 4i\tau_0 \sum_j\omega_j + \tau_0^2 \sum_j\omega_j^2 + 3\tau_0^2 \sum_j\prod_{k\neq j}\omega_k \right) }{ \left(\prod_j(1 + i\tau_0\sum_{k \neq j}\omega_k)\right) \left(1 + i\tau_0\sum_j\omega_j \right) }. \nonumber
\end{align}

\begin{figure*}[t]
    \centering
    \includegraphics[width=1.6\columnwidth]{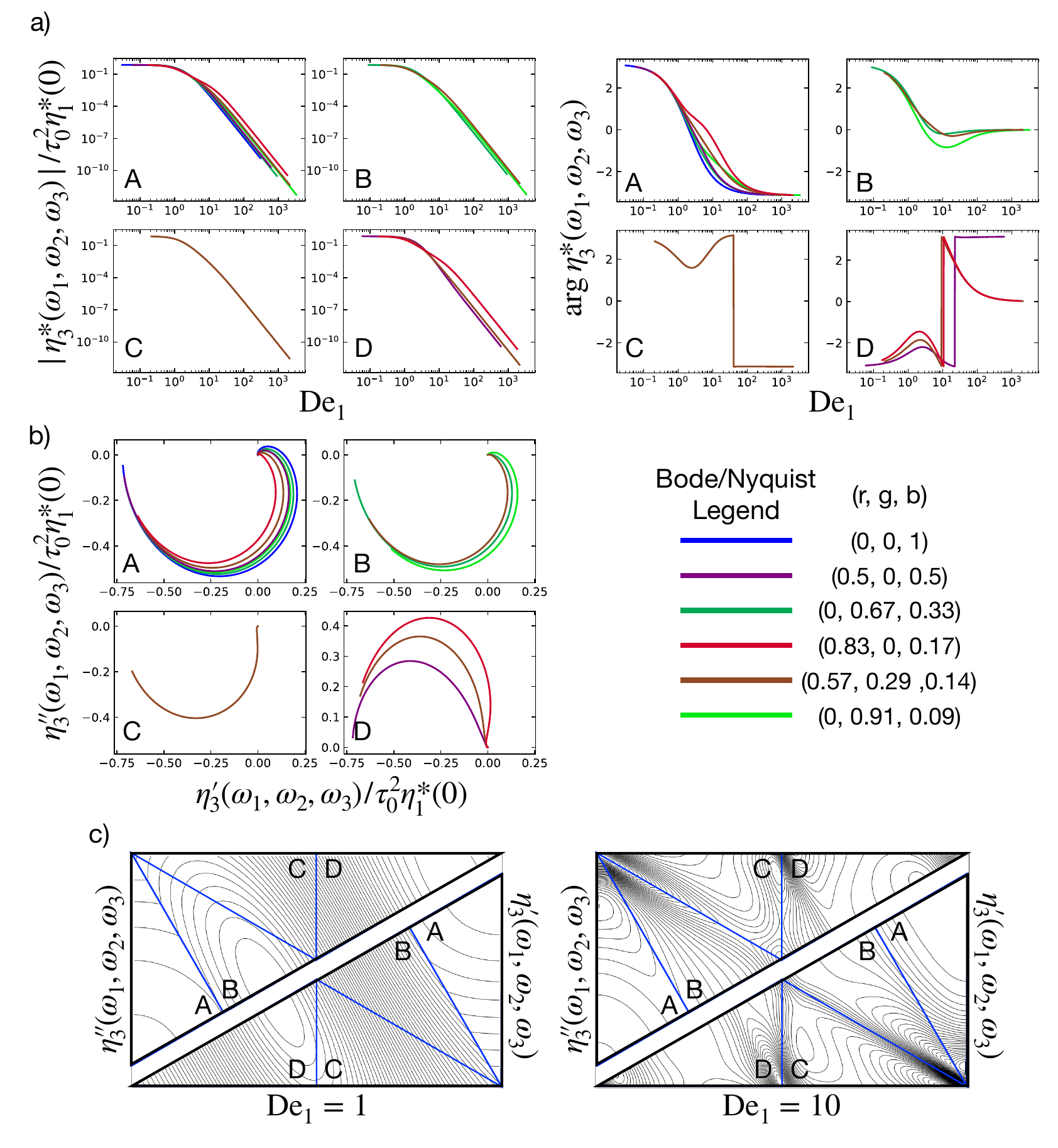}
    \caption{MAPS rheology for a Giesekus fluid with $ \alpha = 0.3 $. a) Bode plots of the magnitude and phase as a function of the $L^1$-norm frequency in different MAPS triangles.  b) Nyquist plots in different MAPS triangles.  c) Contour plots at different $ \mathrm{De}_1 $ on the constant $L^1$-norm surface.}
    \label{fig:giesekus_MAPS}
\end{figure*}

Figure \ref{fig:giesekus_MAPS} presents Bode and Nyquist diagrams along with select contour plots of the Giesekus model with $\alpha = 0.3$. It is clear that the MAPS response of a Giesekus fluid is distinct from all other models considered in this work. Most notably, $|\eta^*_3(\omega_1,\omega_2,\omega_3)|$ scales as $\text{De}_1^{-4}$ at high frequencies, a stronger dependence than observed for the previous models. Considering the above discussion of time-strain-separability, and given that the linear response function for the Giesekus model is of the same single-mode Maxwell form as for the fluidity and single-mode Lodge models, this indicates that the Giesekus model is indeed not time-strain separable, consistent with past observations \cite{ewoldt-2019}. Some other general features of the Giesekus solution, such as the general pattern of the contours at $\text{De}_1 = 1$ and $\text{De}_1 = 10$ and the high-frequency limits of $\text{arg}\eta^*_3(\omega_1,\omega_2,\omega_3)$ being integer multiples of $\pi$, are better captured by the time-strain separable solution presented in Figure \ref{fig:tss} than for the other models. Thus while the time-strain separable solution does not capture the behavior of the Giesekus model in all MAPS deformation histories, in some simple flows we might expect a Giesekus fluid to be approximated well by a time-strain separable model. Larson and Khan, for example, have observed that the response of the Giesekus model during stress relaxation after step-shear closely resembles the expected behavior for a time-strain separable model with damping function $ h( \gamma^2 ) = ( 1 + \alpha( 1-\alpha) \gamma^2 )^{-1} $ \cite{larson-1987}. The Giesekus model exhibits the expected low-frequency expansion for a viscoelastic fluid, with the coefficients:
\begin{equation}
    \left( \begin{array}{c} a \\ b \\ c\\ d \end{array} \right) = \frac{\eta_0 \tau_0^2 \alpha}{3} \left( \begin{array}{c} -9 + 6\alpha \\ i\tau_0 (21 - 16\alpha) \\ \tau_0^2 (59 - 50\alpha) \\  \tau_0^2 (37 - 30\alpha) \end{array} \right). \label{eq:lowfreqgiesekus}
\end{equation}
With a single parameter $\alpha$, there is no way to make this low frequency expansion match the other models. We also see that, because $\alpha < 1$ in the Giesekus model, the parameter $a$ is always negative, consistent with shear thinning behavior.

\section{Discussion}
\label{NLSI}

One goal of experimental rheometry is nonlinear system identification (NLSI).  This is the task of determining an appropriate mathematical model relating the stress response of a material to its deformation history.  NLSI involves four principal operations: data collection, model postulation, parameter estimation, and model validation \cite{nelles-2001}. The tasks of data collection and model postulation,  in particular, are highly specialized and often coupled by the experimental design process.  In these tasks, the choice of a specific experimental technique and corresponding mathematical framework in which data is analyzed has direct and substantial impacts. Many experimental frameworks limit the amount of unique information that can be learned about a material. As we will discuss shortly, this amounts essentially to sampling from low-dimensional manifolds in a high-dimensional response space. Such limitations may have critical impacts on the latter stages of NLSI, such as reducing the number of parameter groups that can be uniquely estimated or making the estimation problem ill-posed. We argue that the MAPS framework is especially well suited for NLSI, both because of the availability of high data-throughput experimental techniques and because of the high dimensionality of its domain. In this section, we discuss the  role of dimensionality in NLSI applied to rheology. However, more detailed examination of experimental MAPS techniques is left for the future, including Part 2 of this work.

Before embarking on an abstract discussion of  the dimensionality of different experimental frameworks, we should consider the steps involved in NLSI in the context of rheology. In rheology, data collection is accomplished by loading a sample in a rheometer, then deforming the material subject to some prescribed protocol and measuring the associated stress response, or stressing the material according to some protocol and recording the deformation history.  For simple shear deformation, a wide variety of different protocols have been developed to aid in the study of nonlinear viscoelasticity, for example: SAOS, MAOS, LAOS, PS, and now MAPS. Each of these protocols will yield one or more data points describing material functions in the corresponding framework, for example $G^*(\omega)$ in SAOS, the intrinsic nonlinear material functions ($[e_1]$, $[v_1]$, $[e_3]$, and $[v_3]$) in MAOS, and $G^*_3(\omega_1,\omega_2,\omega_3)$ or $\eta^*_3(\omega_1,\omega_2,\omega_3)$ in MAPS. Obviously, experimental protocals that yield a larger number of distinct points of data are inherently advantageous in NSLI. An optimally windowed chirp in SAOS, for example, is able to quickly estimate $G^*_1(\omega)$ over a range of $\omega$ \cite{geri-2018}. A simple MAPS experiment, in which an oscillatory shear composed of three superimposed tones excites the weakly nonlinear shear stress response in a viscoelastic material, can measure up to 19 unique complex data points at once, an information-rich data set. The generality of MAPS even permits more complex shear stress or strain signals to be used, which can yield many more distinct measurements of the response function in similar or equal time. Though we leave further discussion of MAPS experiments to Part 2 of this work, it is already clear that such experiments are well suited for the data collection task in NLSI.

Once sufficient data has been collected to characterize a material, the next step in NLSI is to postulate an appropriate constitutive model relating the material's stress and strain histories. Postulating an appropriate model is done independently from the experimental framework used to collect and analyze data, and relies heavily on domain expertise. Once a model has been selected, parameter estimation can, in principle, be performed by numerically evaluating the model equations directly for the selected experimental protocol, and minimizing an appropriately defined loss function between model predictions and experimental results. However, direct numerical evaluation of the model equations can be quite expensive and prone to numerical error; thus model postulation often requires an additional step of obtaining an analytic solution to the model equations for the selected experimental protocol. Such solutions have been obtained for a variety of constitutive models in SAOS, MAOS, and LAOS. Section \ref{sec:examples} presented a few such solutions for MAPS, indicating how these solutions could be used to infer the values of nonlinear model parameters (such as $\alpha$ in the Giesekus model) or distributions (such as $H_1(\tau)$ in the Lodge-like model).

For some materials, it may be possible to postulate a constitutive model that accurately describes the linear and nonlinear relationship between the stress and strain histories. For many other materials of interest, however, this is not the case. One distinctive challenge in NLSI as applied to rheology is the fact that the nonlinear relationship between strain history and stress response has been worked out for only limited classes of material type, flow geometry, and flow history \cite{larson-1999}.  The creation and validation of constitutive equations for particular materials and flows is an active area of research \cite{morris-2009,fischer-2014,larson-2015,beris-2015,geri-2017}.  However, the first principles modeling of novel material formulations that are relevant in industrial applications has proven especially challenging \cite{mewis-2009}.  Even if appropriate models exist for a material under study, it remains an open question whether the data collected from common nonlinear rheological experiments is rich enough to distinguish between different model postulates or to classify complex multi-component formulations.  

For the case of the linear viscoelastic response in equilibrated materials, the above question can be answered affirmatively.  We know from first principles that linear viscoelasticity derives from an underlying distribution of relaxation times associated with microscale relaxation processes in a viscoelastic material \cite{tschoegl-1989,honerkamp-1989}.  In a large number of cases, we understand how to manipulate the molecular composition of a material to achieve a particular relaxation time distribution \cite{graessley-1974,winter-1986,mellema-1989}.  NLSI in linear viscoelasticity, therefore, amounts to inferring this relaxation time distribution from data.  When using data collected from a set of SAOS experiments with the frequency swept across a broad range encompassing the characteristic rates of relaxation in the material, this inference problem is well-posed; thus NLSI applied to linear viscoelasticity is indeed possible, and is widely applied today. To see why this is the case, we revisit the concept of dimensionality and its role in NLSI.

\begin{figure}[t]
    \centering
    \includegraphics[scale=0.6]{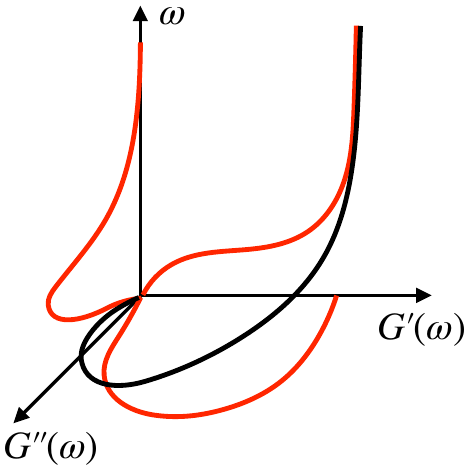}
    \caption{The one dimensional manifold embedded in $ \mathbb{R} \times \mathbb{C} $ measured in a SAOS experiment (black curve) for a viscoelastic fluid material.  A Nyquist plot projects this manifold into the $ G^\prime( \omega ) $ -- $ G^{\prime\prime}( \omega ) $ plane, while a Bode plot projects the manifold into the $ G^\prime( \omega ) $ -- $ \omega $ and $ G^{\prime\prime}( \omega ) $ -- $ \omega $ planes.}
    \label{fig:manifold}
\end{figure}

A SAOS frequency sweep imposes an oscillatory deformation of small strain amplitude at different frequencies, $ \omega \in \mathbb{R} $.  For each frequency probed, a SAOS experiment measures a complex valued linear response function, for example the complex modulus, $ G^*( \omega ) = G^\prime( \omega ) + i G^{\prime\prime}( \omega ) \in \mathbb{C} $. When viewed from a mathematical perspective, the SAOS experiment represents a map from the space of oscillation frequencies into the space of linear response functions: $ f_\mathrm{LR}: \mathbb{R} \rightarrow \mathbb{C} $.  For a particular material, this map describes a one dimensional manifold -- a curve parameterized by the frequency and embedded in the space: $ \mathbb{R} \times \mathbb{C} $ as illustrated in Figure \ref{fig:manifold}.  A typical experimental SAOS frequency sweep samples from this manifold, and the data from such an experiment can be transformed into an approximation of the underlying relaxation time distribution -- another linear manifold \cite{orbey-1991,honerkamp-1989,baumgaertel-1989}.  The unique identification of the linear response via the relaxation time distribution is possible because the collected data sampled from the SAOS manifold is of equal or higher dimension than the manifold underlying the postulated model \cite{graessley-1974,winter-1986}.

For nonlinear viscoelasticity, it is not clear which experiments provide sufficiently rich data for system identification.  In general, we know only a little about the mathematical structure underlying nonlinear viscoelasticity.  For particular classes of materials (e.g. polymer melts or solutions), theories can be developed that are consistent with underlying molecular relaxation mechanisms and descriptive of specific experiments and flow histories, but across different classes of complex fluids and soft materials there are different physical considerations that confound development of universal relations \cite{carreau-1968,acrivos-1973,phillips-1992,sollich-1998}.  Postulates from rational mechanics restrict somewhat the form constitutive models can take \cite{rivlin-1959,rivlin-1971}, but there is not a general relationship between the nonlinear viscoelastic response and some set of physically rationalized degrees of freedom equivalent to the relationship between the linear viscoelastic response and the equilibrium relaxation time distribution in a material.  Empiricism must be employed to close this gap, and this has lead to the development of a variety of different data collection methods that probe nonlinear and time-dependent mechanical responses on higher dimensional manifolds to test these approximations \cite{hyun-2011,dealy-1999}.  One hope is that if data can be drawn from a manifold of sufficiently high dimension, then it may be possible to rationalize the structure of constitutive models and gain insight into the nonlinear response of a particular material formulation. For example, if the same data set is collected across many materials and formulations, it may be possible in the near future to use data-driven methods of classification, such as machine learning, to automatically identify materials with qualitatively similar nonlinear viscoelastic responses.

To further explore these ideas we consider two (closely-related) oscillatory flow protocols used to systematically characterize nonlinear viscoelasticity: MAOS and LAOS.  The MAOS experiment utilizes a single tone oscillatory strain as in SAOS to excite weak nonlinearities in the shear stress. As we discussed in detail in section \ref{sec:maos}, these can be characterized by four real-valued, frequency dependent quantities termed \textit{intrinsic nonlinearities} \cite{ewoldt-2013}:
\begin{equation}
    \left([e_1](\omega), [v_1](\omega), [e_3](\omega), [v_3](\omega) \right) \in \mathbb{R}^4. 
\end{equation}  
The MAOS experiment samples data from the map: $ f_\mathrm{MAOS} : \mathbb{R} \rightarrow \mathbb{R} ^ 4 $. In section \ref{sec:maos}, it was shown that the viscous intrinsic nonlinear functions ($[v_1](\omega)$, $[v_3](\omega)$) compose the real part of the third order complex viscosity along certain rays, and the elastic intrinsic nonlinear functions ($[e_1](\omega)$, $[e_3](\omega)$) compose the imaginary component of the third order complex viscosity along these rays; thus the MAOS experiment's map can be equivalently written as $f_\mathrm{MAOS}$: $\mathbb{R} \rightarrow \mathbb{C}^2$ to closer reflect its similarity to the map represented by SAOS.  The MAOS map describes a one dimensional manifold embedded in the five dimensional space: $ \mathbb{R}^5 $ (or $\mathbb{R} \times \mathbb{C}^2$).  It remains to be established whether it is possible to uniquely identify viscoelastic models via this low dimensional manifold \cite{ewoldt-2018}, though recent work has demonstrated that many constitutive models posses unique signatures in their intrinsic nonlinear functions \cite{bharadwaj-2015}.

The LAOS experiment imposes a single tone strain with high amplitude and measures odd integer harmonics of this tone in the shear stress \cite{hyun-2011}.  A set of these experiments sweeps both the oscillation frequency and the amplitude of deformation, $ \gamma_0 \in \mathbb{R} $, which together compose the well-known Pipkin space \cite{pipkin-1972}.  LAOS probes the map: $ f_\mathrm{LAOS}: \mathbb{R}^2 \rightarrow \mathbb{C}^{N} $, where $ N $ is the number of odd harmonics that can be reliably measured.  The LAOS map describes a two dimensional manifold -- a surface parameterized by the frequency and strain amplitude -- embedded in the high dimensional space: $ \mathbb{R}^2 \times \mathbb{C}^N $.  The LAOS experiment has been referred to by some of the present authors as a rheological fingerprint \cite{mckinley-2008}.  Indeed, it describes a higher dimensional manifold than that sampled in SAOS or MAOS, and may be more suitable to data-driven analysis protocols suggested earlier.  Another seeming advantage of the LAOS experiment is that the LAOS response surface can be lifted into progressively higher dimensions by measuring ever more odd harmonics.  However, this does not change the underlying dimensionality of the LAOS manifold, and accurate measurement of higher harmonics requires painstaking experimentation \cite{hyun-2011}.  It is simply unclear whether such effort is worthwhile in the context of NLSI.  Additionally, for some materials and flow conditions $(\omega, \gamma_0 ) $, LAOS experiments are known to induce shear bands \cite{gurnon-2012} that reflect a multiplicity in the underlying nonlinear response of the material.  NLSI can still be used when the underlying system possesses such features by introducing perturbations that stabilize one particular state \cite{billings-1980}.  However, such methods are yet to be explored in the field of rheology.

In comparison, for a MAPS experiment, a time-periodic shear protocol composed of many distinct tones is used to promote a weakly nonlinear stress response in a viscoelastic material.  This response corresponds  to the MAPS response functions spanning the space: $ ( \omega_1, \omega_2, \omega_3 ) \in \mathbb{R}^3 $. Like the Pipkin space in LAOS, this three frequency space represents the domain over which materials are studied using MAPS. For each frequency combination we measure a single complex-valued nonlinear response function on these three variables denoted $ G_3^*( \omega_1, \omega_2, \omega_3 ) \in \mathbb{C} $.  The MAPS experiment generates data representing the map: $ f_\mathrm{MAPS} : \mathbb{R}^3 \rightarrow \mathbb{C} $, which describes a three dimensional manifold -- a volume parameterized by the three input tones -- embedded in the space: $ \mathbb{R}^3 \times \mathbb{C} $.  By design, the structure of this data set is reminiscent of that resulting from SAOS experiments -- an $ M $ dimensional manifold embedded in the space: $ \mathbb{R}^M \times \mathbb{C} $, with $ M = 1 $ for SAOS and $ M = 3 $ for MAPS.  MAPS data is also sampled from a higher dimensional manifold than either MAOS or LAOS.  The third order response functions are a  rheological realization of ``big data'' that may prove to be more useful than existing methods and measures at enabling future data-driven efforts in the field. We will report on experimental measurements of MAPS data in Part 2 of this work.

\section{Conclusion}

From the above discussion, it is clear that the MAPS framework possesses distinct advantages for NLSI applied to weakly nonlinear rheology as compared to widely applied frameworks such as MAOS and LAOS. Especially for data-driven efforts in NLSI, it is critical to sample from as high-dimensional a space as possible. The three-dimensional domain of MAPS, therefore, is preferable to the two-dimensional domain of LAOS or one-dimensional domain of MAOS. The higher dimensionality of MAPS experiments originates from the generality of the MAPS framework, which is equipped to describe an arbitrary simple shear deformation protocol. Besides its impacts in NLSI, this generality gives MAPS the ability to do something that both MAOS and LAOS cannot: to use information collected by one deformation protocol to predict the weakly linear response to another deformation protocol. Just as it is possible to convert from $G^*(\omega)$, which easily describes oscillatory shear experiments, to $G(t)$, which describes experiments such as step-shear or shear start-up, in the linear viscoelastic regime, knowledge of $G^*_3(\omega_1,\omega_2,\omega_3)$ from multi-tone oscillatory experiments allows the prediction of the incipient nonlinearities in a step-strain or shear start-up experiment, for example. While in another framework it is unclear how to make such generalizations, or if they are possible at all, they are natural using the tools of MAPS rheology.

In this work, the mathematical foundations of MAPS rheology have been developed, including the frequency-domain Volterra representation of nonlinear rheology, important symmetries of the third order material functions, relationships between MAPS and other common experimental protocols, a geometric representation of the MAPS domain, and solutions of simple constitutive models in the MAPS framework. All of these developments are independent of experimental details, such as the choice of specific deformation protocols used to measure the MAPS material functions. Part 2 of this work will revisit the multi-tone input signal discussed in the introduction of this part in the context of experimental protocol design within the MAPS framework. Considering only three-tone inputs, Part 2 will examine features of the experimental design such as (i) practices for choosing appropriate values of the the input harmonics $n_m$ to explore specific areas of the MAPS domain, and (ii) finding the appropriate range for the amplitude of the input signal based on a thorough analysis of errors due to bias resulting from higher-order effects and variance resulting from experimental noise. Part 2 will also detail the procedure for data analysis, including a polynomial regression scheme to isolate third order features of the material response, and how to translate the material response to discrete values of the third order MAPS functions. Finally, real experimental MAPS data for a model complex fluid system -- a surfactant solution of entangled wormlike micelles -- will be presented, along with a comparison to the corotational Maxwell model solution presented in the present Part 1.

Though Part 1 and Part 2 of this work together comprise a detailed study of MAPS rheology, there are still many open questions in this field. For instance, the convergence of the Volterra series cannot be guaranteed for materials without fading memory, such as thixotropic materials, as was briefly discussed in section \ref{volterra}. What input functions should be used to study these types of materials, or what tests can be used to determine whether a specific material response can be completely characterized by a fading memory, are unknown. Further, a quadratic low-frequency expansion of the third order complex viscosity for materials with a longest relaxation time was proposed in section \ref{sec:lowfreq}, without proof that this expansion is valid for all such materials. Though all models studied herein permit this expansion, is it not known whether any material with a longest relaxation time exhibits a third order complex viscosity that is non-analytic before second order in frequency, or how it can be proven that no such material can exist. Additionally, how or whether the MAPS framework is capable of completely describing the behavior of materials that exhibit features such as plasticity or a yield stress, which might not obey some of the fundamental assumptions regarding the symmetries underlying MAPS rheology, has not been explored. All of these questions make for interesting areas of future research.

Another open challenge is in the interpretation of MAPS data. Though some general principles for the interpretation of the MAPS functions have been presented, such as the real and imaginary part of the third order complex viscosity representing viscous and elastic nonlinearities, it is unclear if the MAPS material functions can directly provide more insight to a material's physics. Gaining insight to molecular physics from data, one of the principal objectives of rheological characterization, should therefore be pursued by comparison with constitutive models on a fluid-by-fluid basis. Still, though MAPS provides uniquely high-dimensional data that is a superset of the data obtained by MAOS or medium amplitude PS experiments, it is not possible to guarantee that all aspects of a complex fluid's response can be distinguished at third order; thus it is possible that two distinct materials cannot be uniquely identified via MAPS alone. Consequently, determining the extent to which MAPS data can inform inferences about materials, and how those inferences can be enhanced by other forms of data, will remain a critical question as MAPS rheology is further developed.

\section*{Acknowledgements}

K.R.L. was supported by the US Department of Energy Computational Science Graduate Fellowship program under grant DE-SC0020347. The authors would like to thank M. Gonzalez for helpful discussions.

\bibliography{biblio}



\appendix

\section{Relationships between strain- and stress-controlled Volterra kernels}
\label{app:stressstrain}

As demonstrated in equations \ref{eq:G_to_J} and \ref{eq:eta_to_phi}, the stress-controlled MAPS functions, $J^*_3(\omega_1,\omega_2,\omega_3)$ and $\phi^*_3(\omega_1,\omega_2,\omega_3)$, can be determined directly from the strain or strain-rate controlled MAPS functions $G^*_3(\omega_1,\omega_2,\omega_3)$ and $\eta^*_3(\omega_1,\omega_2,\omega_3)$, with sufficient knowledge of the linear response functions, $G^*_1(\omega)$ and $\eta^*_1(\omega)$. Specifically, the value of a stress-controlled MAPS function at the point $(\omega_1,\omega_2,\omega_3)$ is equal to the negative of the corresponding strain or strain-rate controlled MAPS function at $(\omega_1,\omega_2,\omega_3)$ scaled by the product of the linear response at each frequency coordinate and the linear response at the frequency sum $\omega_1+\omega_2+\omega_3$. A proof of the relationship between $J^*_3(\omega_1,\omega_2,\omega_3)$, $G^*_3(\omega_1,\omega_2,\omega_3)$, and $G^*_1(\omega)$ is shown below. The proof is identical for the relationship between $\phi^*_3$, $\eta^*_3$, and $\eta^*_1$, and can be applied in reverse to relate the strain or strain-rate controlled MAPS functions to the corresponding stress-controlled functions at first and third order.

Consider a MAPS experiment in which both the stress and strain signals can be expressed as a power series in some characteristic amplitude, $\epsilon$, with odd powers only to preserve odd symmetry. To third order, the signals are:
\begin{equation}
    \sigma(t) = \epsilon \sigma^{(1)}(t) + \epsilon^3 \sigma^{(3)}(t) + O(\epsilon^5),
\end{equation}
\begin{equation*}
    \gamma(t) = \epsilon \gamma^{(1)}(t) + \epsilon^3 \gamma^{(3)}(t) + O(\epsilon^5),
\end{equation*}
where $\sigma^{(1)}$ and $\gamma^{(1)}$ represent the linear component of the stress and strain signals, and $\sigma^{(3)}$ and $\gamma^{(3)}$ represent the third order component of the stress and strain signals. With these observations alone, the problem can be formulated as either stress or strain controlled:
\begin{align}
    \hat{\sigma}(\omega) &= G^*_1(\omega)\hat{\gamma}(\omega) \\
    & + \frac{1}{4\pi^2}\iiint_{-\infty}^{\infty}G^*_3(\omega_1,\omega_2,\omega_3)\delta(\omega - \sum_j\omega_j) \nonumber \\
    & \quad \quad \quad \times \hat{\gamma}(\omega_1)\hat{\gamma}(\omega_2)\hat{\gamma}(\omega_3)d\omega_1d\omega_2d\omega_3 + O(\hat{\gamma}^5), \nonumber
\end{align}
\begin{align*}
    \hat{\gamma}(\omega) &= J^*_1(\omega)\hat{\sigma}(\omega) \\
    & + \frac{1}{4\pi^2}\iiint_{-\infty}^{\infty}J^*_3(\omega_1,\omega_2,\omega_3)\delta(\omega - \sum_j\omega_j) \\
    & \quad \quad \quad \times \hat{\sigma}(\omega_1)\hat{\sigma}(\omega_2)\hat{\sigma}(\omega_3)d\omega_1d\omega_2d\omega_3 + O(\hat{\sigma}^5).
\end{align*}
If we substitute the expansion for $\hat{\sigma}$ into the expansion for $\hat{\gamma}$, we find that:
\begin{align}
    \epsilon\hat{\gamma}^{(1)} & + \epsilon^3\hat{\gamma}^{(3)} + O(\epsilon^5) =  \\ & \epsilon J^*_1(\omega)G^*_1(\omega)\hat{\gamma}^{(1)}(\omega) + \epsilon^3 J^*_1(\omega)G^*_1(\omega)\hat{\gamma}^{(3)}(\omega) \nonumber \\
    & + \frac{\epsilon^3 J^*_1(\omega)}{4\pi^2}\iiint_{-\infty}^{\infty}G^*_3(\omega_1,\omega_2,\omega_3)\delta(\omega - \sum_j\omega_j) \nonumber \\
    & \quad \quad \quad \times \hat{\gamma}^{(1)}(\omega_1)\hat{\gamma}^{(1)}(\omega_2)\hat{\gamma}^{(1)}(\omega_3)d\omega_1d\omega_2d\omega_3 \nonumber \\
    & + \frac{\epsilon^3}{4\pi^2} \iiint_{-\infty}^{\infty}J^*_3(\omega_1,\omega_2,\omega_3)G^*_1(\omega_1)G^*_1(\omega_2)G^*_1(\omega_3) \nonumber \\
    & \quad \quad \quad \times \delta(\omega - \sum_j\omega_j)\hat{\gamma}^{(1)}(\omega_1)\hat{\gamma}^{(1)}(\omega_2)\hat{\gamma}^{(1)}(\omega_3) \nonumber \\
    & \quad \quad \quad \times d\omega_1d\omega_2d\omega_3. \nonumber
\end{align}
Equating terms from the left and right hand side of this expression we find that at $O(\epsilon)$:
\begin{equation}
    \hat{\gamma}^{(1)} = J^*_1(\omega)G^*_1(\omega)\hat{\gamma}^{(1)},
\end{equation}
which gives the familiar relationship between the Fourier transforms of the linear response functions (i.e. the complex modulus and the complex compliance):
\begin{equation}
    J^*_1(\omega) = \frac{1}{G^*_1(\omega)}.
    \label{eq:linear_relationship}
\end{equation}
At $O(\epsilon^3)$, the expression can be simplified using equation \ref{eq:linear_relationship} to give:
\begin{align}
    0 &= \\
    & \frac{1}{G^*_1(\omega)}\iiint_{-\infty}^{\infty}G^*_3(\omega_1,\omega_2,\omega_3)\delta(\omega - \sum_j\omega_j) \nonumber \\
    & \quad \quad \quad \times \hat{\gamma}^{(1)}(\omega_1)\hat{\gamma}^{(1)}(\omega_2)\hat{\gamma}^{(1)}(\omega_3)d\omega_1d\omega_2d\omega_3 \nonumber \\
    & + \iiint_{-\infty}^{\infty}J^*_3(\omega_1,\omega_2,\omega_3)G^*_1(\omega_1)G^*_1(\omega_2)G^*_1(\omega_3) \nonumber \\
    & \quad \quad \quad \times \delta(\omega - \sum_j\omega_j)\hat{\gamma}^{(1)}(\omega_1)\hat{\gamma}^{(1)}(\omega_2)\hat{\gamma}^{(1)}(\omega_3) \nonumber \\
    & \quad \quad \quad \times d\omega_1d\omega_2d\omega_3. \nonumber
\end{align}
The sifting property of the delta function allows $G^*_1(\omega)$ to be moved inside the first integral by the variable substitution $\omega \rightarrow \omega_1+\omega_2+\omega_3$, which gives:
\begin{align}
    & - \iiint_{-\infty}^{\infty}\frac{G^*_3(\omega_1,\omega_2,\omega_3)}{G^*_1(\omega_1+\omega_2+\omega_3)}\delta(\omega - \sum_j\omega_j) \\
    & \quad \quad \quad \times \hat{\gamma}^{(1)}(\omega_1)\hat{\gamma}^{(1)}(\omega_2)\hat{\gamma}^{(1)}(\omega_3)d\omega_1d\omega_2d\omega_3 \nonumber \\
    & = \iiint_{-\infty}^{\infty}J^*_3(\omega_1,\omega_2,\omega_3)G^*_1(\omega_1)G^*_1(\omega_2)G^*_1(\omega_3) \nonumber \\
    & \quad \quad \quad \times \delta(\omega - \sum_j\omega_j)\hat{\gamma}^{(1)}(\omega_1)\hat{\gamma}^{(1)}(\omega_2)\hat{\gamma}^{(1)}(\omega_3) \nonumber \\
    & \quad \quad \quad \times d\omega_1d\omega_2d\omega_3. \nonumber
\end{align}
By comparison of the integrands, therefore, we find that:
\begin{align}
    & J^*_3(\omega_1,\omega_2,\omega_3) = \nonumber \\
    & \quad -\frac{G^*_3(\omega_1,\omega_2,\omega_3)}{G^*_1(\omega_1)G^*_1(\omega_2)G^*_1(\omega_3)G^*_1(\omega_1+\omega_2+\omega_3)}.
\end{align}

\section{Relating MAOStrain and MAOStress}
\label{app:maostress}

In Section \ref{sec:maos}, we use the definition of MAOS in strain control (MAOStrain) to derive relationships between the third order complex modulus or viscosity and the four intrinsic nonlinear functions in MAOStrain: $[e_1](\omega)$, $[v_1](\omega)$, $[e_3](\omega)$, and $[v_3](\omega)$. There exists a similar formulation of MAOStress \cite{ewoldt-2013}. If the stress is imposed as a single tone oscillation, $\sigma = \sigma_0\cos(\omega t)$, then the medium-amplitude shear strain response can be expanded as a cubic polynomial in the stress amplitude:
\begin{align}
    \gamma(t) &= \sigma_0\left[J'(\omega)\cos(\omega t) + J''(\omega)\sin(\omega t)\right] \\
    &+ \sigma_0^3 \Big([c_1](\omega)\cos(\omega t) + \frac{1}{\omega}[f_1](\omega)\sin(\omega t) \nonumber \\
    &+ [c_3](\omega)\cos(3\omega t) + \frac{1}{3\omega}[f_3](\omega)\sin(3\omega t)\Big) \nonumber.
\end{align}
Where $[c_n]$ and $[f_n]$ are called the intrinsic compliance coefficients and intrinsic fluidity coefficients, respectively. Using a procedure analogous to that described in Section \ref{sec:maos}, we find relationships between these intrinsic nonlinear functions and the third order complex modulus or fluidity:
\begin{subequations}
\label{eq:maostress_all}
\begin{align}
    &[c_1](\omega) = \frac{3}{4}J'_3(\omega,-\omega,\omega) = \frac{3}{4\omega}\phi''_3(\omega,-\omega,\omega), \\
    &[f_1](\omega) = \frac{3\omega}{4}J''_3(\omega,-\omega,\omega) = \frac{3}{4}\phi'(\omega,-\omega,\omega), \\
    &[c_3](\omega) = \frac{1}{4}J'_3(\omega,\omega,\omega) = \frac{1}{12\omega}\phi''_3(\omega,\omega,\omega), \\
    &[f_3](\omega) = \frac{3\omega}{4}J''_3(\omega,\omega,\omega) = \frac{1}{4}\phi'_3(\omega,\omega,\omega).
\end{align}
\end{subequations}
Using either of the inversion relationships, equation \ref{eq:G_to_J} or \ref{eq:eta_to_phi}, along with the MAOS relationship in equations \ref{eq:maos_all} and \ref{eq:maostress_all}, it is possible to obtain relationships between the MAOStrain and MAOStress functions. Derivation of these relationships is algebraically tedious, so we omit the details here. The resulting MAOS inversion relationships are:
\begin{subequations}
\begin{align}
    &[c_1](\omega) = -\frac{R_1(\omega)[e_1](\omega) + \omega I_1(\omega)[v_1](\omega)}{R_1^2(\omega) + I_1^2(\omega)}, \\
    &[f_1](\omega) = \frac{\omega\left(\omega R_1(\omega) [v_1](\omega) - I_1(\omega)[e_1](\omega)\right)}{R_1^2(\omega) + I_1^2(\omega)},  \\
    &[c_3](\omega) = \frac{\omega I_3(\omega)[v_3](\omega) - R_3(\omega)[e_3](\omega)}{R_3^2(\omega) + I_3^2(\omega)}, \\
    &[f_3](\omega) = -\frac{3\omega\left(I_3(\omega)[e_3](\omega) + \omega R_3(\omega) [v_3](\omega)\right)}{R_3^2(\omega) + I_3^2(\omega)},
\end{align}
\end{subequations}
with:
\begin{equation}
    R_1(\omega) = G'(\omega)^4 - G''(\omega)^4,
\end{equation}
\begin{equation}
    I_1(\omega) = 2G'(\omega)G''(\omega)\left[G'(\omega)^2 + G''(\omega)^2\right] ,\nonumber
\end{equation}
\begin{align}
    R_3(\omega) =& G'(\omega)^3 G'(3\omega) - 3G'(\omega)G''(\omega)^2 G'(3\omega) \nonumber \\
    &- 3G'(\omega)^2 G''(\omega) G''(3\omega) + G''(\omega)^3 G''(3\omega), \nonumber
\end{align}
\begin{align}
    I_3(\omega) =& 3G'(\omega)^2 G''(\omega)G'(3\omega) - G''(\omega)^3 G'(3\omega) \nonumber \\
    &+ G'(\omega)^3 G''(3\omega) - 3G'(\omega) G''(\omega)^2 G''(3\omega). \nonumber
\end{align}

\section{Derivation of $\eta^*_3(\omega_1,\omega_2,\omega_3)$ for the Corotational Maxwell Model}
\label{app:crm}

The corotational Maxwell model is another tensorial differential model of widespread use. It can be expressed by the differential equation:
\begin{equation}
    \boldsymbol{\sigma} + \tau_0\frac{\mathcal{D}\boldsymbol{\sigma}}{\mathcal{D}t} = \eta_0\boldsymbol{\dot{\gamma}},
\end{equation}
where $\frac{\mathcal{D}}{\mathcal{D}t}$ represents the \textit{corotational} derivative of the stress tensor:
\begin{equation}
    \frac{\mathcal{D}\boldsymbol{\sigma}}{\mathcal{D}t} \equiv \frac{D\boldsymbol{\sigma}}{Dt} + \frac{1}{2}\left\{\boldsymbol{\omega}\cdot\boldsymbol{\sigma} - \boldsymbol{\sigma}\cdot\boldsymbol{\omega}\right\},
\end{equation}
with the vorticity tensor:
\begin{equation}
    \boldsymbol{\omega} \equiv \nabla\textbf{u} - (\nabla\textbf{u})^T, \nonumber
\end{equation}
and the deformation protocol $\textbf{u} = \gamma_0s(t)x_2\textbf{e}_1$. To find the form of the third order complex viscosity, we expand the stress tensor as a power series in the characteristic amplitude $\gamma_0$:
\begin{equation}
    \boldsymbol{\sigma} = \gamma_0\boldsymbol{\sigma}^{(1)} + \gamma_0^2\boldsymbol{\sigma}^{(2)} + \gamma_0^3\boldsymbol{\sigma}^{(3)} + O(\gamma_0^4). \label{eq:power_series}
\end{equation}
Terms at $O(\gamma_0^4)$ and above are not considered in the medium-amplitude limit. The contributions to the shear stress at each order are found by asymptotic matching. At $O(\gamma_0)$, we find:
\begin{equation}
    \sigma_{12}^{(1)} + \tau_0\frac{d\sigma_{12}^{(1)}}{dt} = \eta_0s(t), \nonumber
\end{equation}
with $\sigma_{11}^{(1)} = \sigma_{22}^{(1)} = 0$. The Fourier transform allows us to determine the linear response function for the corotational Maxwell model model,
\begin{equation}
    \eta^*_1(\omega) = \frac{\eta_0}{1 + i\omega\tau_0}.
    \label{eq:maxwell_eta1}
\end{equation}
This is the familiar linear response function for a single Maxwell mode.

At $O(\gamma_0^2)$, we find that:
\begin{equation}
    \sigma_{11}^{(2)} + \tau_0\frac{d\sigma_{11}^{(2)}}{dt} - \tau_0s(t)\sigma_{12}^{(1)} = 0,
\end{equation}
\begin{equation}
    \sigma_{22}^{(2)} + \tau_0\frac{d\sigma_{22}^{(2)}}{dt} + \tau_0s(t)\sigma_{12}^{(1)} = 0,
\end{equation}
and $\sigma_{12}^{(2)} = 0$. Finally, at $O(\gamma_0^3)$:
\begin{equation}
    \sigma_{12}^{(3)} + \tau_0\frac{d\sigma_{12}^{(3)}}{dt} + \frac{1}{2}\tau_0\left(\sigma_{11}^{(2)} - \sigma_{22}^{(2)}\right) = 0,
\end{equation}
and $\sigma_{11}^{(3)} = \sigma_{22}^{(3)} = 0$. By taking the Fourier transforms of the above differential equations and substituting in the expressions for $\hat{\sigma}_{12}^{(1)}(\omega)$, $\hat{\sigma}_{11}^{(2)}(\omega)$, and $\hat{\sigma}_{22}^{(2)}(\omega)$ into that for $\hat{\sigma}_{12}^{(3)}(\omega)$, we are able to obtain an expression for $\hat{\sigma}_{12}^{(3)}(\omega)$ in terms of $\omega$ and the model parameters only. Obtaining this expression in closed form requires application of the convolution theorem, resulting in convolutions of the Fourier-transformed functions of the form:
\begin{equation}
    a(\omega)\{b(\omega) * [c(\omega)(d(\omega) * e(\omega))]\}.
\end{equation}
Such expressions can be rewritten in integral form as:
\begin{align}
     & a(\omega)\{b(\omega) * [c(\omega)(d(\omega) * e(\omega))]\} = \nonumber \\
     & \iiint_{-\infty}^{\infty}a(\omega_1+\omega_2+\omega_3)b(\omega_1)c(\omega_2+\omega_3)d(\omega_2)e(\omega_3) \nonumber \\
     & \quad\quad\quad\quad \times \delta(\omega - \omega_1 - \omega_2 - \omega_3)d\omega_1d\omega_2d\omega_3 \label{eq:conv_identity}.
\end{align}
\vfill\null
This expression is of the form of the third order Volterra integral. However, it does not obey permutation symmetry with respect to the arguments $(\omega_1,\omega_2,\omega_3)$. This can be easily fixed as done for other models by averaging the contribution from each permutation of the arguments. By applying this identity to the closed form expression for the Fourier transform of $\sigma_{12}^{(3)}$, we can obtain the expression for $\eta^*_3(\omega_1,\omega_2,\omega_3)$:
\begin{align}
    & \frac{\eta^*_3(\omega_1,\omega_2,\omega_3)}{\eta_0\tau_0^2} = -\frac{1}{6}\left(\frac{1}{1 + i\tau_0\sum_j\omega_j}\right) \label{eq:crm_eta3} \\
    & \times \sum_j \left[\left(\frac{1}{1 + i\tau_0\sum_{k\neq j}\omega_k}\right)\sum_{k\neq j}\left(\frac{1}{1 + i\tau_0\omega_k}\right)\right]. \nonumber
\end{align}

As previously noted, the corotational Maxwell model is an example of a time-strain separable model \cite{ewoldt-2019}. If we substitute the expression for the first order complex viscosity, given in \ref{eq:maxwell_eta1}, into equations \ref{eq:G_to_eta} and subsequently into \ref{eq:tss}, we find that a fluid that obeys time-strain separability and displays a linear response characterized by \ref{eq:maxwell_eta1} will have a third order complex viscosity of the form:
\begin{align}
    & \frac{\eta^*_3(\omega_1,\omega_2,\omega_3)}{\eta_0\tau_0^2} = \left.\frac{\partial h(\gamma^2)}{\partial(\gamma^2)}\right\rvert_{\gamma=0}\left(\frac{1}{1 + i\tau_0\sum_j\omega_j}\right) \label{eq:tss_eta3} \\
    & \times \sum_j \left[\left(\frac{1}{1 + i\tau_0\sum_{k\neq j}\omega_k}\right)\sum_{k\neq j}\left(\frac{1}{1 + i\tau_0\omega_k}\right)\right]. \nonumber
\end{align}
Comparison of equations \ref{eq:crm_eta3} and \ref{eq:tss_eta3} reveal that the corotational Maxwell model indeed is consistent with time-strain separability, with the condition that:
\begin{equation}
    \left.\frac{\partial h(\gamma^2)}{\partial(\gamma^2)}\right\rvert_{\gamma=0} = -\frac{1}{6}.
\end{equation}

This result is consistent with that presented by Martinetti and Ewoldt in MAOS \cite{ewoldt-2019}. In fact, it is straightforward to show that \ref{eq:tss_eta3} reduces to equation 28a and 28c of reference \cite{ewoldt-2019} when the substitution $(\omega_1,\omega_2,\omega_3) \xrightarrow{} (\omega,\omega,-\omega)$ is made, and to equation 28b and 28d of reference \cite{ewoldt-2019} when the substitution $(\omega_1,\omega_2,\omega_3) \xrightarrow{} (\omega,\omega,\omega)$ is made.

\section{Derivation of $\eta^*_3(\omega_1, \omega_2, \omega_3 )$ for the Giesekus model}
\label{app:giesekus}

The Giesekus constitutive model is a tensorial differential model represented by the equation:
\begin{equation}
    \boldsymbol{\sigma} + \tau_0\boldsymbol{\sigma}_{(1)} + \frac{\alpha\tau_0}{\eta_0}\boldsymbol{\sigma}\cdot\boldsymbol{\sigma} = \eta_0\boldsymbol{\dot{\gamma}},
\end{equation}
where $\boldsymbol{\sigma}_{(1)}$ represents the upper convected derivative of the stress tensor:
\begin{equation}
    \boldsymbol{\sigma}_{(1)} = \frac{D \boldsymbol{\sigma}}{D t} - (\nabla\textbf{u})^T \cdot \boldsymbol{\sigma} - \boldsymbol{\sigma} \cdot (\nabla\textbf{u})
\end{equation}
with the rate-of-strain tensor:
\begin{equation}
    \boldsymbol{\dot{\gamma}} = \nabla\textbf{u} + (\nabla\textbf{u})^T.
\end{equation}
In the Giesekus model, $\boldsymbol{\sigma}$ represents the polymer contribution to the stress tensor, which is often superimposed on a Newtonian solvent contribution. In MAPS rheology, the deformation protocol is $\textbf{u} = \gamma_0 s(t) x_2 \textbf{e}_1$, where $\gamma_0$ is some characteristic amplitude of the deformation, and $s(t)$ is an arbitrary function of time.

To find the form of the third order complex viscosity for the Giesekus model, we expand the stress tensor as a power series as in equation \ref{eq:power_series} and perform asymptotic matching. At $O(\gamma_0)$, we find that:
\begin{equation}
    \sigma_{12}^{(1)} + \tau_0\frac{d\sigma_{12}^{(1)}}{dt} = \eta_0 s(t),
\end{equation}
and that $\sigma_{11}^{(1)} = \sigma_{22}^{(1)} = 0$. We see that this is the same differential equation as for the corotational Maxwell model, thus the first order complex viscosity for the Giesekus model is given by equation \ref{eq:maxwell_eta1}.

At $O(\gamma_0^2)$, we find that:
\begin{equation}
    \sigma_{11}^{(2)} + \tau_0\frac{d\sigma_{11}^{(2)}}{dt} - 2\tau_0 s(t) \sigma_{12}^{(1)} + \frac{\alpha \tau_0}{\eta_0}\left(\sigma_{12}^{(1)}\right)^2 = 0,
\end{equation}
\begin{equation}
    \sigma_{22}^{(2)} + \tau_0\frac{d\sigma_{22}^{(2)}}{dt} + \frac{\alpha \tau_0}{\eta_0}\left(\sigma_{12}^{(1)}\right)^2 = 0,
\end{equation}
and $\sigma_{12}^{(2)} = 0$. At $O(\gamma_0^3)$, we find that
\begin{equation}
    \sigma_{12}^{(3)} + \tau_0 \frac{d\sigma_{12}^{(3)}}{dt} - \tau_0 s(t) \sigma_{22}^{(2)} + \frac{\alpha \tau_0}{\eta_0}\sigma_{12}^{(1)}\left(\sigma_{11}^{(2)} + \sigma_{22}^{(2)}\right) = 0.
\end{equation}

By taking the Fourier transform of the differential equations for $\sigma_{11}^{(2)}$, $\sigma_{22}^{(2)}$, and $\sigma_{12}^{(3)}$, substituting in the known form of the Fourier transform of $\sigma_{12}^{(2)}$, we can obtain an expression for $\sigma_{12}^{(3)}$ in terms of $\omega$ and the model parameters only. Using the convolution identity presented in equation \ref{eq:conv_identity}, we can obtain the expression for $\eta^*_3(\omega_1,\omega_2,\omega_3)$:
\begin{align}
    & \frac{ \eta^*_3( \omega_1, \omega_2, \omega_3 ) }{ \eta_0 \tau_0^2 } = \frac{ \alpha \left( (3 - 2\alpha) + i\tau_0 \sum_j\omega_j \right) }{ 3 \left(\prod_j(1 + i\tau_0\omega_j)\right) } \times \label{eq:giesekus_eta3_app} \\
    & \frac{ \left(-3 - 4i\tau_0 \sum_j\omega_j + \tau_0^2 \sum_j\omega_j^2 + 3\tau_0^2 \sum_j\prod_{k\neq j}\omega_k \right) }{ \left(\prod_j(1 + i\tau_0\sum_{k \neq j}\omega_k)\right) \left(1 + i\tau_0\sum_j\omega_j \right) }. \nonumber
\end{align}

\end{document}